\documentclass[a4paper, 11pt]{article}

\usepackage[font={small,it}]{caption}
\usepackage{graphicx}
\usepackage{mathtools}
\usepackage{xcolor}
\usepackage{slashed}
\usepackage{amsmath}
\usepackage{amssymb }
\usepackage{mathtools}
\usepackage{caption}
\usepackage{latexsym}
\usepackage{float}
\usepackage{braket}
\usepackage{wrapfig}
\usepackage{tabularx}
\usepackage{textgreek}
\usepackage{verbatim}
\usepackage{physics}
\usepackage{subfigure}
\usepackage[utf8]{inputenc}
\usepackage{geometry}
\definecolor{MyDarkBlue}{rgb}{0.15,0.15,0.45}
\usepackage[linktocpage=true]{hyperref}
\hypersetup{
colorlinks=true,
citecolor=MyDarkBlue,
linkcolor=MyDarkBlue,
urlcolor=MyDarkBlue,
}

\usepackage[numbers,sort&compress]{natbib}

\date{}

\begin{document}
\def\thefootnote{\fnsymbol{footnote}}
\begin{center}

{\LARGE
\textbf{Gravitational Backreaction in de Sitter:\\
A Canonical ADM Approach}
\par}

\vspace{0.5cm}

{\large
Giordano Cintia\footnote{giordano.cintia@univ-amu.fr}
\par}

\vspace{0.5cm}

{\small
\textit{Aix Marseille Universit\'e, Universit\'e de Toulon, CNRS, CPT, Marseille, France}
\par}

\vspace{0.2cm}

\end{center}

\vspace{.6cm}

\hrule \vspace{0.2cm}
\centerline{\small{\bf Abstract}}
\vspace{0.1cm}
{\small\noindent 
We develop a canonical ADM framework for gravitational backreaction during inflation. Our approach extends the formalism used to compute cosmological correlators by applying it to the evolution of the quantum background, which we identify with the metric one-point function. The aim is to assess the quantum stability of inflationary geometries under gravitational backreaction. Within this framework, the backreaction is determined by the expectation values of the Hamiltonian constraint and the equations of motion for the spatial metric. Working in the exact de Sitter limit, we apply the framework to a massless minimally coupled spectator scalar field and the physical graviton polarizations. At one loop, their backreaction renormalizes the relation between the cosmological constant and the Hubble parameter without inducing any secular departure from de Sitter evolution, in agreement with previous results. We perform the calculation using a hard momentum cutoff and show that a fixed physical cutoff allows the homogeneous background equations to be renormalized with time- and background-independent coefficients, in contrast to a fixed comoving cutoff. Finally, we demonstrate that the explicit representation of the backreaction depends on the gauge condition imposed on the metric fluctuations, as well as on the background–fluctuation split. Nevertheless, these different representations are related by a gauge transformation and are therefore physically equivalent.
}

\vspace{0.3cm}
\noindent
\hrule
\def\thefootnote{\arabic{footnote}}
\setcounter{footnote}{0}
\newpage
\tableofcontents
\newpage
\section{Introduction}

Inflation provides the leading framework for explaining the origin of cosmic structure: microscopic quantum fluctuations generated during this accelerated phase are stretched to cosmological scales and become the seeds of the density perturbations observed in the Cosmic Microwave Background and in the large-scale structure of the Universe. As cosmological data continue to improve our sensitivity to primordial fluctuations, understanding the quantum origin of these observables becomes increasingly important for connecting the inflationary theory with observations~\cite{Weinberg:2005vy, Weinberg:2006ac, Senatore:2009cf, Pimentel:2012tw, Green:2020whw}.  
Most inflationary predictions are computed in a rigid, or fixed-background, approximation, in which quantum fluctuations propagate on the classical geometry. In natural units, this approximation is formally obtained by taking $H/M_{\rm pl}\to0$, with the Hubble scale $H$ kept fixed. 
However, quantum fluctuations also carry energy and momentum, and therefore backreact on the metric. Although this gravitational backreaction is expected to be subleading in many inflationary settings, it could become relevant when (and if) small effects have a sufficiently long time to accumulate.\footnote{A significant amount of work has also focused on the backreaction of quantum fluctuations on the inflaton dynamics, in regimes where the spacetime geometry is still treated as fixed~\cite{Boyanovsky:2004ph,Boyanovsky:2005sh, Boyanovsky:2005px}. See also~\cite{ Franciolini:2023agm, Kristiano:2022maq, Inomata:2025bqw, Braglia:2025cee, Braglia:2025qrb, Fumagalli:2023zzl,Fumagalli:2024jzz, Caravano:2024moy, Fang:2025vhi, Inomata:2025pqa, Braglia:2026fzp} and references therein for recent applications in the context of ultra slow-roll inflation. }  

Due to the quasi-de Sitter nature of the inflationary background, it is considered to be useful to understand this phenomenon in the exact de Sitter limit, where the dynamics of the inflaton background trivializes.  The graviton itself is, however, part of the degrees of freedom of the system and backreacts on the evolution of the background. This raises a long-standing but basic question: are de Sitter solutions stable under radiative corrections? 

In this paper, we develop a framework for addressing this question within the canonical ADM formulation of General Relativity. The main novelty is that, within this formalism, gravitational backreaction is encoded directly in the dynamics of the metric one-point function. This makes it possible to track explicitly how the same quantum background geometry is represented under different choices of background–fluctuation split, graviton parametrization, and gauge fixing for the background and fluctuation sectors.

Our approach is guided by the idea of~\cite{Dvali:2013eja} that, in a fundamentally quantum theory, a classical inflationary geometry should emerge from a quantum state possessing an appropriate degree of classicality, rather than being introduced as a fundamentally classical background. In inflationary models whose potential vanishes at its minimum, the underlying theory admits a Minkowski vacuum, which provides a natural reference state for this construction. The time-dependent inflationary geometry can then be viewed as arising from an excited semiclassical state—such as a coherent state with some non-Gaussian corrections required for consistency~\cite{Berezhiani:2023uwt}—built over this vacuum and evolved unitarily. See~\cite{Glauber:1963tx,Kibble:1965zza,Zhang:1990fy,Zhang:1999is} for foundational work on coherent states, Refs.~\cite{Berezhiani:2020pbv,Berezhiani:2021gph,Berezhiani:2025tkp} for recent applications to the dynamics of scalar backgrounds, and Refs.~\cite{Berezhiani:2021zst,Berezhiani:2024boz,Berezhiani:2025roh} for coherent-state constructions of gravitational backgrounds within canonical BRST quantization; see also~\cite{Berezhiani:2024pub}. Throughout this work, we regard exact de Sitter space as a limiting case of such an inflationary spacetime, rather than as the vacuum of a theory with a fundamental positive cosmological constant. 

Motivated by this perspective, we begin with the full, unshifted ADM variables, before introducing any split between background and fluctuations. Following the discussions of~\cite{Berezhiani:2021gph,Berezhiani:2023uwt, Berezhiani:2025tkp}, if such a state exists, the time evolution of these operators can be reorganized as a semiclassical expansion around their one-point functions in the semiclassical state, rather than around a prescribed solution of the classical equations of motion. The quantum-corrected geometry is then identified with the metric one-point function, whose evolution is determined self-consistently together with that of the higher-order matter and metric correlation functions evaluated in the same state. The conventional background and its quantum fluctuations thus emerge as different correlation functions of the same underlying quantum degrees of freedom, rather than as independently defined ingredients. In the present work, we do not attempt to construct this semiclassical state explicitly. Instead, we characterize it operationally through the initial conditions imposed on correlation functions.

There are several motivations for introducing such a framework. First, cosmological correlation functions are commonly computed within the canonical ADM formalism, with the transverse-traceless gauge imposed on metric fluctuations~\cite{Arnowitt:1962hi, Maldacena:2002vr, Cheung:2007st, Piazza:2013coa}. Formulating gravitational backreaction in the same language makes it possible to connect this problem directly with the technically mature methods developed for computing inflationary correlators.

Moreover, claims of de Sitter instability have also emerged from complementary particle-based descriptions of gravitational backgrounds, providing an additional motivation for such a framework. The first indication is provided by the aforementioned corpuscular approach~\cite{Dvali:2013eja}. From this perspective, identifying a classical gravitational configuration with a stable vacuum may be misleading: interactions among its quantum constituents can gradually drive the background away from its classical trajectory and erode its coherence, invalidating the classical description after a characteristic timescale known as the \emph{quantum break-time}~\cite{Dvali:2013vxa}. Such effects have been investigated for black holes~\cite{Dvali:2012en,Dvali:2013eja}, inflationary backgrounds~\cite{Dvali:2013eja, Berezhiani:2016grw,Berezhiani:2015ola,Berezhiani:2022gnv}, and, most relevant to the present work, de Sitter spacetime~\cite{Dvali:2017eba,Dvali:2024dzf}. Related phenomena in nongravitational quantum field theories have been discussed in~\cite{Dvali:2013vxa,Dvali:2017ruz,Dvali:2022vzz}.

A second, related indication comes from the particle-production perspective developed in~\cite{Polyakov:2007mm,Polyakov:2009nq,Polyakov:2012uc}. These works argue that particle production in interacting quantum field theories on de Sitter spacetime may destabilize the quantum background, with infrared effects potentially modifying its global structure. This possibility will reappear below in the conventional field-theoretic analysis of de Sitter correlation functions. The underlying physical picture is analogous to the depletion of a classical electric field through Schwinger pair production.
Both approaches suggest that the classical description of a de Sitter background may break down after a characteristic timescale as a result of the cumulative effects of quantum interactions. It is therefore desirable to formulate and quantify these effects using standard quantum field theory and nonequilibrium techniques.

The paper is structured as follows. 
In Section 2, we review the relation between infrared divergences and the possible quantum breaking of de Sitter isometries, and explain why semiclassical calculations may appear to indicate an instability of the de Sitter background. In Section 3, we reformulate the problem of computing gravitational backreaction in the ADM framework. We apply this formulation to derive the one-loop backreaction from a massless spectator scalar field and the physical graviton helicities.
We work in a cutoff regularization scheme and show that a cutoff on physical modes provides counterterms with background-independent coefficients, in contrast to a cutoff on comoving modes. Since hard cutoffs break diffeomorphism invariance, power-law divergences must be canceled by nongeometric counterterms chosen so that the renormalized background equations satisfy the Bianchi identities. We will show that the one-loop gravitational backreaction of a minimally coupled massless free scalar and the
graviton amounts to a shift of the Hubble parameter, without generating a physical secular instability
of the background, in agreement with results obtained by functional methods~\cite{Miao:2017vly}.
In Section 4, we examine how different gauge-fixing choices for the fluctuations induce different slicings of the quantum-corrected background. In Section 5, we discuss possible extensions of the framework to higher loop orders. Finally, in Section 6, we summarize our results.
\paragraph{\textit{Notation and conventions}}
We work in natural units, $\hbar=c=1$, and set $M^{-2}_\text{pl}=8\pi G_N=1$ unless otherwise specified. We adopt the mostly-plus signature $(-,+,+,+)$. Unless specified otherwise, all field operators entering a composite operator are understood to depend on the same spacetime coordinates $(x,t)$. Throughout, we omit hats on operators.

\section{Overview of Gravitational Backreaction in de Sitter}
\label{sec2}

Several infrared properties of quantum fields in de Sitter have been interpreted as possible indications of a quantum instability of the background. For example, it has long been known that a massless minimally coupled scalar field quantized on a de Sitter spacetime does not admit a perturbative de Sitter-invariant Fock vacuum~\cite{Allen:1985ux}. Since in the transverse-traceless (TT) gauge the infrared behavior of the graviton is closely related to that of such a scalar~\cite{Antoniadis:1986sb}, one might suspect that tensor fluctuations backreact on the geometry in a way that spoils de Sitter invariance. Perturbatively, the non-invariance of the vacuum state of fluctuations appears as secular growth in coincident two-point functions, sourced by an infrared divergence in the theory. At the tree level and in TT gauge, this reads
\begin{equation}
\langle \gamma_{ij}\gamma^{ij}\rangle = \frac{\Lambda^2_{\rm UV}}{\pi^2 M^2_\text{pl}}+\frac{2H^2}{M^2_\text{pl}\pi^2}\left(\log \Lambda_{\rm UV} L+Ht\right),
\label{eq:quadraticIntro}
\end{equation}
where $\Lambda_{\rm UV} $ and $L^{-1}$ are respectively the physical UV and comoving IR cutoffs. In the above, we have also temporarily restored $M_{\rm pl}$.

Nevertheless, explicit computations of gravitational backreaction induced by gravitons lead to the more conservative conclusion that the de Sitter isometries are preserved at one loop~\cite{Tsamis:1996qq, Tsamis:1996qm,Miao:2017vly}. The reason is the following: consider the effective energy-momentum tensor of cosmological perturbations given in~\cite{Mukhanov:1996ak, Abramo:1997hu}. When applied to tensor fluctuations, the corresponding quadratic contribution can be interpreted as the backreaction of gravitons on the background geometry. In the de Sitter limit, the quantities that define the quadratic energy-momentum tensor are only derivative correlators. At tree level, such derivative correlators do not inherit the same secular growth as in Eq.~\eqref{eq:quadraticIntro}, as follows from differentiating the non-coincident two-point function before taking the coincidence limit. Thus, at this order, the backreaction renormalizes the local parameters of the gravitational action without producing a secular departure from de Sitter geometry~\cite{Birrell:1982ix}.  Similar  conclusions have been obtained in Ref.~\cite{Janssen:2008dw}.

The situation becomes more subtle at higher loop order. Graviton-mediated corrections to the same derivative correlators can generate secular contributions~\cite{Giddings:2010nc}. Such terms may then enter the effective energy-momentum tensor, and have been argued to induce secular evolution of the background~\cite{Tsamis:1996qq, Tsamis:1996qm}.\footnote{A similar mechanism was proposed in the context of interacting scalar field theories, such as a massless scalar field with quartic interactions~\cite{Ford:1984hs}, without the need to account for gravitons. However, for such theories, the
leading secular dynamics can be resummed with the formalism of stochastic inflation~\cite{Starobinsky:1986fx,Starobinsky:1994bd}, while
subleading secularities can be treated systematically within field theory approaches~\cite{Gorbenko:2019rza, Baumgart:2019clc}.}
The main subtlety concerns the relation between the gauge choice used for the fluctuations and the gauge in which the quantum-corrected background is described. In Ref.~\cite{Tsamis:1996qq}, the computation for backreaction is performed by splitting the conformal background from fluctuations, which are then quantized in the BRST formalism with a de Donder-type gauge fixing. This setup complicates the interpretation of backreaction because the gauge-fixing condition imposed on the fluctuations is not, in general, compatible with keeping the corrected metric in the same conformal FLRW slicing, as we discuss in Section~4. In other words, the backreaction appears in a form compatible with the prescribed de Donder gauge imposed on the fluctuations, rather than necessarily preserving the conformal FLRW slicing of the original prescribed background.  The resulting apparent de Sitter-breaking backreaction may therefore be difficult to disentangle from the change of slicing induced by the interplay between the fluctuation gauge and the background--fluctuation split. One possible solution would be to construct gauge-invariant observables~\cite{Garriga:2007zk}, which is the approach adopted in Ref.~\cite{Miao:2017vly} for the one-loop computation. In this case, the mismatch between the gauge choice for the background and that for the fluctuations would not be problematic in itself, since the result would be gauge invariant. A second approach, which is the one we pursue in this paper, is to look for a gauge choice for fluctuations that is compatible with a quantum-corrected FLRW cosmological slicing.\footnote{It has also been argued that, in certain gauges for the fluctuations, the infrared pathology of the graviton propagator is absent~\cite{Allen:1986ta, Higuchi:2011vw}, and a de Sitter-invariant graviton two-point function can be constructed. However, since a change of gauge for the fluctuations can affect the background–fluctuation split beyond linear order, one should still assess the role of backreaction within these gauges. Thus, a major caveat in this class of computations is whether any de Sitter-breaking contribution is physical or merely a gauge artifact.}

A second, more technical subtlety concerns the ultraviolet regularization of cosmological correlators. If a sharp momentum cutoff is employed within an effective field theory description, it is natural to impose it at a fixed physical scale. Since loop integrals are expressed in terms of comoving momenta, this corresponds to a time-dependent comoving cutoff.  In Sect.~\ref{sec:cutoff}, we show that a fixed physical cutoff allows the homogeneous background equations to be renormalized with background-independent, time-independent counterterm coefficients. By contrast, a fixed comoving cutoff would require explicitly time-dependent and background-dependent coefficients.

Taken together, these issues motivate a framework in which the background–fluctuation split, gauge fixing, and regulator dependence can all be tracked explicitly.

\section{{ADM Framework for Gravitational Backreaction}}

In this section, we consider the ADM formulation of General Relativity, with the goal of deriving the background dynamics from the expectation values of the constraint and evolution equations. 
We consider a free massless scalar field $\psi$ minimally coupled to gravity
\begin{equation}
S=\frac{1}{2}\int {\rm{d}}^4x \sqrt{-g}\,\biggr\{R-g^{\mu \nu}\partial_\mu\psi \partial_\nu \psi-2\Lambda_{\rm cc}\biggr\},
\label{eq:action}
\end{equation}
where $R$ is the scalar curvature and $\Lambda_{\rm cc}$ is the cosmological constant. The scalar field is introduced as a useful comparison with respect to metric perturbations. 

\subsection{ADM formulation}
The first obstacle to quantizing this theory is the gauge redundancy of General Relativity.  In the ADM formulation, the spacetime metric $g_{\mu\nu}$ is decomposed into variables that make the constraint structure explicit, thereby separating the dynamical spatial metric from the nondynamical lapse function and shift vector. 

To follow this approach, we introduce the following line element
\begin{equation}
ds^2=-(N^2-N^iN_i)dt^2+2N_i dtdx^i+h_{ij}dx^i dx^j,
\label{eq:ADMline}
\end{equation}
which is the ADM line element. In this expression, $N$, $N^i$, and $h_{ij}$ are the lapse function, the shift vector, and the spatial metric, respectively. Up to a total derivative, the action reads
\begin{equation}
S=\frac{1}{2}\int d^4x \sqrt{h}\left\{N{R}^{(3)}+\frac{1}{N}\left(E_{ij}E^{ij}-E^2\right)+\frac{1}{N}\left(\dot{\psi}-N^i\partial_i\psi\right)^2-Nh^{ij}\partial_i\psi\partial_j\psi- 2N{\Lambda_\text{cc}} \right\}.
\label{eq:ADMaction}
\end{equation}
Here, $E_{ij}$ reads
\begin{equation}
E_{ij}=\frac{1}{2}\left\{\dot{h}_{ij}-h_{ik}\partial_jN^k-h_{jk}\partial_i N^k-N^k\partial_kh_{ij}\right\},
\end{equation}
while $R^{(3)}$ is the spatial curvature and $h=\det \,h_{ij}$. Within perturbation theory, one may fix a gauge and solve the constraint equations that we give below, order by order, for the nondynamical lapse and shift. The resulting reduced phase space, comprising the physical gravitational and matter degrees of freedom, can then be perturbatively quantized by imposing the standard equal-time canonical commutation relations. A semiclassical state representing the inflationary geometry may subsequently be selected within the associated Hilbert space.

In fact, the advantage of the ADM action \eqref{eq:ADMaction} is that the lapse $N$ and the shift $N^i$ appear as Lagrange multipliers. Their equations of motion impose the Hamiltonian and momentum constraints, respectively, given by
\begin{flalign}
&{R}^{(3)}-N^{-2}\left(E_{ij}E^{ij}-E^2\right)-N^{-2}\left(\dot{\psi}-N^i\partial_i\psi\right)^2-h^{ij}\partial_i\psi\partial_j\psi-2{\Lambda_\text{cc}}=0~,\label{eq:lapseEq} \\
&\nabla_j\left[N^{-1}\left(E_i^j-\delta_i^j E\right)\right]-N^{-1}\left(\dot{\psi}-N^j\partial_j\psi\right)\partial_i\psi =0\label{eq:shiftEq}.
\end{flalign}
The time evolution of the spatial metric is described by the highly non-linear equation for $h_{ij}$
\begin{flalign}
\ddot{h}_{ij}+\frac{1}{2}\dot{h}_{km}h^{km}\dot{h}_{ij}-\dot{h}_{ik}&h^{km}\dot{h}_{mj}+2N^2R^{(3)}_{ij}=2N^2{\Lambda_\text{cc}} h_{ij}+2N^2\partial_i\psi\partial_j\psi,
\label{eq:spatial}
\end{flalign}
where the operator $R^{(3)}_{ij}$ is the three-dimensional Ricci tensor. Here, we have omitted terms involving the shift vector, derivatives of the lapse function, and terms that are quadratic in these non-dynamical variables. The rationale is that such contributions do not affect the one-loop background equations considered below.\footnote{See Appendix~\ref{sec:NandNjeq} for the perturbative solutions of the constraint equations that justify this truncation, and Refs.~\cite{Seery:2006vu,Dimastrogiovanni:2008af} for the corresponding results in standard semiclassical parametrizations.}

This set of equations describes the gravitational sector of the theory, provided that we also include the equation of motion for the scalar field
\begin{equation}
\ddot{\psi}+\partial_t\left(\log\sqrt{h}\right)\dot{\psi}-N^2h^{ij}\partial_i\partial_j\psi-N^2h^{ij}\partial_i\left(\log\sqrt{h}\right) \, \partial_j\psi- N^2\partial_ih^{ij} \partial_j \psi=0,
\end{equation}
where the same truncation has been applied.

\subsection{Formulating a de Sitter state through initial conditions}

The system of equations presented above must be supplemented with the initial conditions specified on a given spacelike hypersurface $t=t_0$.
In quantum field theory, the physical information about a system is encoded in its quantum state. More precisely, a state specified on a spacelike hypersurface $t=t_0$ characterizes all correlation functions that can be constructed from field operators and their conjugate momenta.

Conversely, specifying all such correlation functions on the same hypersurface is sufficient, at least in principle, to reconstruct the quantum state.
However, due to the constraints, only a reduced set of independent initial data needs to be specified in the case above.
Thus, we characterize the de Sitter state $\lvert dS\rangle$ by specifying the initial correlation functions of the independent dynamical fields and their conjugate momenta. The constraint equations instead determine the residual initial data.

For the one-point functions of these operators, we choose initial conditions such that their expectation values give de Sitter in FLRW coordinates in the cosmological slicing. In particular, we impose
\begin{equation}
 \langle {N}(t_0,x)\rangle=1,\quad \langle {N}^j(t_0,x)\rangle=0, \quad \langle {h}_{ij}(t_0,x)\rangle= e^{2\rho_0}\delta_{ij},\quad \langle{\psi}(t_0,x)\rangle=0~.
\label{eq:shift}
\end{equation}
and
\begin{equation}
\langle\dot{N}(t_0,x)\rangle=0, \qquad \langle\dot{N}^j(t_0,x)\rangle=0~,\qquad \langle \dot{h}_{ij}(t_0,x)\rangle=2\dot{\rho}_0 e^{2\rho_0} \delta_{ij}~.
\end{equation}
 The initial value of $\rho_0$ is a free choice, since the scale factor is defined only up to an overall multiplicative normalization. 
The initial expansion rate, by contrast, is not an independent condition. The value of $\dot{\rho}_0$ is fixed by the expectation value of the Hamiltonian constraint and therefore already includes the backreaction of quantum fluctuations. In other words, at the level of the one-point function, we work with the convention that the initial condition on the derivative of the spatial metric—or equivalently the initial expansion rate $\dot\rho_0$—is fixed by the constraints, while the initial conditions for the lapse function and shift vector are given.

It is convenient to introduce the notation 
\begin{equation}
\boxed{
\begin{array}{l}
    \text{One-point}:\qquad \qquad \exp(\rho_0),\qquad \dot{\rho}_0 \\[1mm]
    \text{Classical}:\qquad \qquad \ \ \, \quad a(t_0),\qquad \!H
\end{array}
}
\end{equation}
 The first set refers to the scale factor and expansion parameter built from the one-point function of the metric, in this case on the initial spacelike hypersurface. The second set, denoted by the standard notation and obtained in the formal $\hbar\to0$ limit, refers to their classical counterparts.

The conditions \eqref{eq:shift} do not completely specify the state of the system, since we still have the freedom to choose the initial conditions for higher-order correlation functions involving different combinations of field operators. To specify them, it is convenient to introduce a new set of operators through the relations
\begin{flalign}
&{N}(x,t)=1+{n}(x,t)\label{eq:tadN},\\ &{N}^i(x,t)=n^i=\partial^i{\chi}(x,t)+{v}^i\left(x,t\right)\label{eq:tadNi},\\ &{h}_{ij}(x,t)=e^{2\rho(t)}\left(\delta_{ij}+\gamma_{ij}(x,t)\right)\label{eq:tadg},
\end{flalign}
where the shift $N^i$ has been decomposed into its scalar and divergence-free parts.

This field redefinition is exact, provided that the newly introduced operators satisfy the \emph{tadpole conditions}~\cite{Boyanovsky:1994me}
\begin{equation}
 \langle 
{n}(\vec{x},t)
\rangle=\langle 
 \partial^i\chi(\vec{x},t)\rangle=\langle \gamma_{ij}(\vec{x},t)\rangle=\langle v^i(\vec{x},t)\rangle =0,
\label{eq:tadpole_condition}
\end{equation}
which are imposed at all orders in perturbation theory. The transverse-traceless condition
\begin{equation}
\gamma^i_i=0,\qquad \partial_i\gamma_{ij}=0,
\end{equation}
is imposed on $\gamma_{ij}$.
We emphasize that the expansion is performed around the one-point function of the spatial metric. If the expansion were instead made around another quantity, such as the classical solution, then imposing vanishing expectation values for the fluctuation fields would no longer be possible in general. We return to this point in Sect.~\ref{sect:classical}.

Having introduced these new operators, we require the quadratic and higher-order correlators in $|{\rm dS}\rangle$ to coincide with those of the standard Bunch-Davies vacuum for the fluctuation fields~\cite{Bunch:1978yq}. In particular
\begin{flalign}
\langle {\psi}(t_0,x){\psi}(t_0,y)\rangle &= \langle \Omega_{BD}| {\psi}(t_0,x){\psi}\, (t_0,y)|\Omega_{BD}\rangle,\\
\langle{h}_{ij}(t_0,x){h}_{km}(t_0,y)\rangle &=e^{4\rho_0}\delta_{ij}\delta_{km}+e^{4\rho_0}\langle \Omega_{BD}| {\gamma}_{ij}(t_0,x){\gamma}_{km}\, (t_0,y)|\Omega_{BD}\rangle.
\end{flalign}
The same prescription applies to higher-order correlation functions. Here, $|\Omega_{\rm BD}\rangle$ denotes the \emph{interacting} Bunch–Davies vacuum, whose correlation functions therefore
include corrections beyond the Gaussian approximation\footnote{In other words, $|\Omega_{BD}\rangle=U_I(t_0,-\infty_-)|0_{BD}\rangle$, where $U_I$ is the interaction picture evolution operator, or Møller operator, and $|0_{BD}\rangle$ is the free Bunch-Davies Gaussian vacuum. The time $t_0$ remains a finite reference time at which the initial data for the background are specified.}. Thus, the initial conditions for the original ADM variables and for the fluctuation fields defined around the metric one-point function are related by the field redefinition above. Notice that, for both a massless minimally coupled scalar field and the graviton, the Bunch–Davies prescription must be supplemented by the infrared cutoff $L$.

Concerning the non-dynamical variables $n$, $\chi$, and $v^i$, their initial correlation functions are not independent, but are determined by the constraint equations on the initial spacelike hypersurface $t=t_0$ in terms of those of the dynamical fields $\psi$ and $\gamma_{ij}$, using, for example, the solutions in Appendix~\ref{sec:NandNjeq}. This is the fluctuation-level counterpart of the construction described above for the background. Here, however, one specifies the initial higher-order correlation functions of the spatial metric, while the constraints determine the corresponding initial data for the lapse and shift fluctuations.

The way in which the constraint equations relate the initial conditions also makes the explicit construction of the state non-trivial. Since $N$ and $N^i$ do not possess independent conjugate momenta or standard canonical commutation relations, the state cannot be represented as a displacement operator acting on the Bunch--Davies vacuum in the same way as one would do in ordinary non-constrained quantum field theories~\cite{Berezhiani:2020pbv,Berezhiani:2021gph,Berezhiani:2023uwt,Berezhiani:2025tkp}. A genuine coherent-state construction is therefore more naturally implemented in a BRST quantization, where all degrees of freedom are made dynamical by introducing ghost and auxiliary sectors~\cite{Berezhiani:2021zst, Berezhiani:2024boz}.

\subsection{The one-loop Friedmann equations}
\label{Sect:1loop}
To derive the backreaction, we expand the constraint and evolution equations around the metric one-point function and its fluctuations. Since we are in the cosmic slicing of FLRW, we have
\begin{equation}
    \langle {N}(x,t)\rangle=1,\qquad 
    \langle {N}^i(x,t)\rangle=0,\qquad t>t_0 .
    \label{eq:initialN}
\end{equation}
and
\begin{equation}
    \langle {h}_{ij}(x,t)\rangle=e^{2\rho(t)}\delta_{ij}\,.
\end{equation}
Thus, the lapse function and shift vector one-point functions are fixed by the choice of background slicing, while the dynamics of the quantum-corrected background is encoded in $\rho(t)$, which at this point is an unknown function. It is determined by the expectation values of the Hamiltonian constraint and of the evolution equation for the spatial metric, as we now show.

We derive the equations of motion for the metric one-point function. We substitute the field redefinitions into the Hamiltonian constraint \eqref{eq:lapseEq}, the momentum constraint \eqref{eq:shiftEq}, and the equation of motion for $h_{ij}$~\eqref{eq:spatial}. We then expand the resulting equations to linear order in the non-dynamical variables and to quadratic order in the dynamical degrees of freedom.
The resulting system separates into two sectors: a set of c-number equations governing the background dynamics, and a set of operator-valued equations describing the evolution of the fluctuations.

The c-number equations are obtained by taking the expectation value of the expanded equations in $|{\rm dS}\rangle$. They read
\begin{flalign}
    &\dot{\rho}^2=\frac{\Lambda_{\rm cc}}{3}+\frac{1}{6}\langle\dot{\psi}^2+e^{-2\rho}\partial_i\psi\partial^i \psi\rangle+ \frac{1}{24}\langle \dot{\gamma}_{ij}\dot{\gamma}^{ij}+e^{-2\rho}\partial_k\gamma_{ij}\partial^k\gamma^{ij}+4\dot{\rho}\,\partial_t\left(\gamma_{ij}\gamma^{ij}\right)\rangle+\ldots ~,
    \label{eq:1stFriedmann1loop}\\
    &\left(\ddot{\rho}+3\dot{\rho}^2\right)=\Lambda_{\rm cc}
 +\frac{1}{3}e^{-2\rho}\langle\partial_i\psi\partial^i\psi\rangle+{\frac{1}{12}\langle e^{-2\rho}\partial_k\gamma_{ij}\partial^k\gamma^{ij}+\partial^2_t (\gamma_{ij}\gamma^{ij})+6\dot{\rho}\partial_t(\gamma_{ij}\gamma^{ij})\rangle}+\ldots~,
 \label{eq:2F}\\
 &\langle \dot{\psi}\partial_j\psi+\frac{1}{2}{\gamma}_{mi}\partial_i\dot{\gamma}_{mj}-\frac{1}{2}\partial_j\left({\gamma}_{mi}\dot{\gamma}_{im}\right)\rangle=0 ~.
\label{eq:momentumbck}
\end{flalign}
To obtain Eq.~\eqref{eq:2F}, we trace the equation of motion for $h_{ij}$ as provided by Eq.~\eqref{eq:spatial} and take the expectation value\footnote{The trace is taken with respect to the inverse metric $h^{ij}$}.
For a homogeneous and isotropic state, the expectation value of the momentum constraint, Eq.~\eqref{eq:momentumbck}, vanishes identically and does not provide an independent background equation.

 This set of equations is exact at order $\hbar$, with ellipses representing higher-than-quadratic correlation functions, which contribute at order $\hbar^2$. These equations govern the dynamics of the quantum background, identified with the metric one-point function, and form the lowest-order sector of the Schwinger--Dyson hierarchy.   
 The system is overcomplete, since the Friedmann equations are not independent once the fluctuation equations of motion are imposed. This redundancy provides a useful consistency condition, which will play an important role in Sect.~\ref{sec:cutoff} in renormalizing the background equations.

 In the first and second equations, all non-Hermitian operators have been made Hermitian by symmetrization, e.g. $\langle \dot{\gamma}_{ij}\gamma^{ij}\rangle\to\frac{1}{2}\partial_t\langle \gamma_{ij}\gamma^{ij}\rangle$.  In particular, terms in the extrinsic curvature contain products of metric variables and their time derivatives, whose ordering becomes relevant upon quantization of the fluctuations. This symmetric ordering ensures that the equations determining $\rho$ are real. Without a Hermitian ordering, the expectation value of the metric could acquire an imaginary part, which is inconsistent with the metric being Hermitian. The same prescription applies to the composite operators generated by substituting the constraint solutions back into the action, which, however, play no role in this paper. Notice that the ordering problem associated with implementing the gravitational constraint algebra at the operator level~\cite{DeWitt1967} does not arise directly in this approach, since quantization in TT gauge proceeds by solving the classical constraints in a perturbative expansion and then quantizing the remaining physical degrees of freedom. 

To evaluate the above equations, we have to derive the correlation functions for the fluctuations. 
The equations of motion of the fluctuation fields are obtained by subtracting the equations \eqref{eq:1stFriedmann1loop}, \eqref{eq:2F}, \eqref{eq:momentumbck} from the corresponding shifted operator equations. Since we are interested in the one-loop dynamics, it is sufficient to truncate this system at linear order in the fluctuations. The resulting equations read
 \begin{flalign}
 &{n}=0~,\\
 &{n}^j=0~,\\
 &\ddot{\gamma}_{ij}+3\dot{\rho}\dot{\gamma}_{ij}-e^{-2\rho}\partial_k\partial^k \gamma_{ij}+2\left(\ddot{\rho}+3\dot{\rho}^2-\Lambda_{\rm cc}\right)\gamma_{ij}+ \ldots=0~.
\label{eq:MSgrav}
 \end{flalign}
 
 The Hamiltonian and momentum constraints show that the lapse and shift fluctuations vanish at linear order. This is consistent with the standard solution of the constraint equations on a de Sitter background, where the lapse and shift start at quadratic order in the dynamical fluctuations~\cite{Maldacena:2002vr,Seery:2006vu,Dimastrogiovanni:2008af}. In Appendix~\ref{sec:NandNjeq}, we provide the corresponding solutions when quadratic terms are retained.
Thus, at one loop, the only nontrivial equation in the metric-fluctuation sector is the tensor equation. This is the standard equation of motion for tensor modes in de Sitter. The term in parentheses is the tree-level second Friedmann equation, implying that tensor modes remain gapless at order $\hbar$, as expected. The ellipses denote terms that are quadratic in the fluctuations. These terms generate loop corrections to the graviton propagator and therefore start contributing only at order $\hbar^2$.

The last equation we need is the equation of motion for the spectator scalar field. With the same approximations used in the metric sector, we find
\begin{equation}
\ddot{\psi}+3\dot{\rho}\dot{\psi}-e^{-2\rho}\partial_k\partial^k \psi+\ldots=0~.
\label{eq:1dynPsi}
\end{equation}

\subsection{One-loop Backreaction in de Sitter}
\label{sec:one-loopFr}
The one-loop dynamics of the system is found by evaluating the one-loop Friedmann equations, combined with the linearized equations of motion for the graviton and the scalar field. Since the right-hand side of the background equations is already of order $\hbar$, the fluctuation correlators entering them can be evaluated on the classical de Sitter background, $\dot{\rho}=H$ and $e^{\rho}=a$. 

Thus, the system of equations we have to solve reads
\begin{flalign}
&\dot{\rho}_\text{1-loop}^2=\frac{\Lambda_{\rm cc}}{3}+\frac{1}{6}\langle\dot{\psi}^2+a^{-2}\partial_i\psi\partial^i \psi\rangle+\frac{1}{24}\langle \dot{\gamma}_{ij}\dot{\gamma}^{ij}+a^{-2}\partial_k\gamma_{ij}\partial^k\gamma^{ij}+4 H\partial_t\left({\gamma}_{ij}\gamma^{ij}\right)\rangle~,\label{eq:1Fr1} \\
&\left(\ddot{\rho}_\text{1-loop}+3\dot{\rho}_\text{1-loop}^2\right)=\Lambda_{\rm cc}+\frac{1}{3}a^{-2}\langle\partial_i\psi\partial^i\psi\rangle+\frac{1}{12}\langle a^{-2}\partial_k\gamma_{ij}\partial^k\gamma^{ij}+\partial^2_t (\gamma_{ij}\gamma^{ij})+6H\partial_t(\gamma_{ij}\gamma^{ij})\rangle\label{eq:2Fr1}~,
\end{flalign}
together with the fluctuation equations
\begin{flalign}
&\ddot{\gamma}_{ij}+3H\dot{\gamma}_{ij}-a^{-2}\partial_k\partial^k \gamma_{ij}=0 \label{eq:gamma2}~,\\
&\ddot{\psi}+3H\dot{\psi}-a^{-2}\partial_k\partial^k \psi=0 \label{eq:psi2}~.
\end{flalign}
To evaluate this system, we first solve equations \eqref{eq:gamma2} and \eqref{eq:psi2}. This can be done by defining the appropriate ladder expansion~\cite{Giddings:2010nc}
\begin{flalign}
&{\psi}(x,t)=\int\frac{d^3k}{(2\pi)^{3/2}}\left(b_k \psi_k e^{i k\cdot x}+b_k^\dag \psi_k^* e^{-i k\cdot x}\right)~,\\  &{\gamma}_{ij}(x,t)=\sqrt{2}\sum_{s=+,\times}  \int\frac{d^3k}{(2\pi)^{3/2}}\left(c^s_k \epsilon_{ij}^s \gamma_k e^{i k\cdot x}+{c^s_k}^\dag {\epsilon^*_{ij}}^s \gamma_k^* e^{-i k\cdot x}\right)~,
\end{flalign}
where the square root of two in front of tensor modes ensures that the polarizations are canonically normalized.
Here, $[b_k,b^\dagger_{k'}]=\delta^{(3)}(k-k')$ and $[c^s_k,{c^{s'}_{k'}}^\dagger]=\delta_{s s'}\delta^{(3)}(k-k')$ 

Also, $\epsilon_{ij}$ is the polarization tensor, satisfying $\epsilon_{ii}=k^i\epsilon_{ij}=0$ and $\epsilon_{ij}^{s}\epsilon_{ij}^{s'}=2\delta_{ss'}$.
The state $|{\rm{dS}}\rangle$ we have chosen implies that we fix mode functions by imposing Bunch--Davies initial conditions, which leads to the following mode functions for the scalar and graviton fields
\begin{equation}
\psi_k(\tau)=\gamma_k(\tau)=-\frac{i H e^{-i k\tau}}{\sqrt{2 k^3}}\left(1+i k \tau\right).
\label{bunch}
\end{equation}
Here, $\tau$ is the conformal time, defined as $dt=\left({-H\tau}\right)^{-1}d\tau$. 

A technical point deserves some care. To evaluate correlation functions, momentum integrals must be regulated by imposing a fixed physical ultraviolet cutoff. In comoving variables, this corresponds to a time-dependent cutoff,
\begin{equation}
    \Lambda_{\rm c}(t)=a(t)\Lambda_{\rm{UV}}~,
\end{equation}
as is customary in cosmological perturbation theory~\cite{Senatore:2009cf, Boyanovsky:2005sh} and as we discuss in the following section. Because the cutoff depends on time, time derivatives do not commute with the momentum integral. Equivalently, differentiating the mode expansion generates boundary terms from the time-dependent upper limit.  It is therefore convenient to rewrite correlation functions involving time derivatives in terms of total time derivatives and spatial derivatives. For instance, focusing on $\langle\dot{\psi}^2\rangle$ and using the equation of motion for the field operator, one obtains\footnote{Recall that the equal-time canonical commutation relation between $\psi$ and $\dot{\psi}$, at this order, reads $[\psi(x),\dot{\psi}(y)]=ia^{-3} \delta^{(3)}(x-y)$. } 
\begin{equation}
    \langle \dot{\psi}^2\rangle=a^{-2}\langle \partial_k\psi\partial^k\psi\rangle+\frac{3}{2}H \partial_t\langle \psi^2\rangle+\frac{1}{2}\partial_t^2 \langle \psi^2\rangle= \frac{1}{16 \pi^2}\left({6 H^4+2 H^2 \Lambda_{\rm{UV}} ^2+\Lambda_{\rm{UV}} ^4}\right),
    \label{eq:psid}
\end{equation}
where we focused on $\hbar$ terms. In other words, correlators involving time derivatives of the fields are reconstructed in terms of total time derivatives of the coincident two-point function and spatial-derivative correlators.
A similar relation holds for tensor modes, using the relation $\langle \dot{\gamma}_{ij}\dot{\gamma}^{ij}\rangle=8\langle \dot{\psi}^2\rangle$.

By substituting these field expansions into the first and second Friedmann equations \eqref{eq:1Fr1} and \eqref{eq:2Fr1} 
we derive the one-loop Hubble parameter
\begin{flalign}
&\dot{\rho}^2_\text{1-loop}=\frac{\Lambda_{\rm cc}}{3}+\frac{1}{48 \pi^2 }\left(3\Lambda_{\rm{UV}}^4+6 H^2\Lambda_{\rm{UV}}^2+25 H^4\right), \label{eq:1fr}\\
&(\ddot{\rho}_\text{1-loop}+3\dot{\rho}_\text{1-loop}^2)=\Lambda_{\rm cc}+\frac{1}{48 \pi^2}\left({3\Lambda_{\rm{UV}} ^4+6 H^2 \Lambda_{\rm{UV}} ^2+48 H^4}\right)
\label{eq:2fr}
\end{flalign}
As we see, the one-loop Hubble parameter inferred from the first Friedmann equation is time-independent. 
\subsection{Renormalization and the choice of UV Cutoff }\label{sec:cutoff}

In this section, we renormalize the one-loop background equations and examine how the procedure depends on whether a fixed ultraviolet cutoff is imposed in physical or comoving momentum.

We begin with the fixed physical UV cutoff. Although the unrenormalized one-loop first and second Friedmann equations are time-independent, the resulting system is not consistent because the solution inferred from the first Friedmann equation does not satisfy the second. To better understand the issue, let us separate the scalar and graviton contributions:
\begin{flalign}
&\dot{\rho}_\text{1-loop}^2=\frac{\Lambda_{\rm cc}}{3}+\frac{1}{48\pi^2 } \left(\Lambda_{\rm UV} ^4+ 2\text{H}^2 \Lambda ^2_{\rm UV}+3H^4\right)_\psi +\frac{1}{24 \pi^2}\left({\Lambda_{\rm UV} ^4}+2 H^2 \Lambda ^2_{\rm UV}+11 H^4\right)_\gamma~, \\
&\left(\ddot{\rho}_\text{1-loop}+3\dot{\rho}^2_\text{1-loop}\right)=\Lambda_{\rm cc} +\frac{1}{48 \pi ^2 }\left(\Lambda_{\rm UV} ^4+ 2\text{H}^2 \Lambda_{\rm UV} ^2\right)_\psi+\frac{1}{24 \pi ^2}\left(\Lambda_{\rm UV} ^4+2 H^2 \Lambda_{\rm UV} ^2+24 H^4\right)_\gamma~.
\label{eq:2Fdiv}
\end{flalign}
 Since scalar and tensor fluctuations enter additively at one loop, each sector must be renormalizable independently. 
 
In a diffeomorphism-invariant scheme, the first and second Friedmann equations are related by the Bianchi identities, or equivalently by the Slavnov--Taylor identities of the quantum theory applied to the one-point function. The structure above shows that, with the cutoff prescription used here, these relations are not preserved. As a result, the divergences cannot be absorbed solely by the standard covariant geometric counterterms. The origin of this breaking lies in the cutoff regularization, which fails to preserve the symmetries of the theory. This is typical of introducing a hard cutoff in gauge theories, which generally requires symmetry-restoring counterterms that are not invariant. The simplest example is QED, where a naive hard cutoff violates the Ward identity in the photon vacuum polarization unless an appropriate symmetry-restoring counterterm is introduced~\cite{Peskin:1995ev}. This violation appears as a power-law cutoff-sensitive local photon mass\footnote{A similar effect arises in the calculation of loop corrections to the graviton two-point function in de Sitter space in Ref.~\cite{Ballesteros:2024cef}. Appendix B of that work also illustrates the role of symmetry-restoring counterterms using vacuum polarization in QED as an example. }.

The way out is to restore covariance at the level of the background equations by introducing counterterms that are not themselves invariant.
The minimal counterterm structure to absorb background divergences in this cutoff scheme is
\begin{equation}
\begin{aligned}
\mathcal{L}_{\text{c.c}}
=N\sqrt{h}\bigg[
-\delta\Lambda_{\rm cc}
+&\frac{\delta M}{2}N^{-2}(E_{ij}E^{ij}-E^2)
+\frac{\alpha}{2}N^2
\\
&+\frac{\beta}{2}(E_{ij}E^{ij}-E^2)
+\frac{\sigma}{2}N^{-2}(E_{ij}E^{ij}-E^2)^2
\bigg].
\end{aligned}
\label{eq:Lcc}
\end{equation}

We can therefore distinguish two classes of counterterms. The counterterms $\delta\Lambda_{\rm cc}$ and $\delta M$ are the usual covariant ones, associated respectively with the vacuum energy and Newton’s constant. Their finite parts are free parameters of the renormalization prescription, reflecting the freedom to fix the corresponding couplings of the theory by measurements at a given renormalization scale.

By contrast, the counterterms proportional to $\alpha$, $\beta$, and $\sigma$ are not independent physical parameters. They encode the symmetry-breaking effects introduced by the cutoff regulator and are fixed by the requirement that the renormalized background equations satisfy the Bianchi identities. Once the covariant counterterms have been chosen, the coefficients of these non-manifestly diffeomorphism-invariant structures are therefore determined by consistency. They should be understood as regulator-dependent restoring counterterms, rather than as new renormalization conditions.


Let us stress that, on homogeneous configurations with vanishing shift vector, a similar
basis of counterterms can be introduced in four-dimensional notation as
\begin{equation}
S=\frac{1}{2}\int d^4x\,\sqrt{-g}
\left(
-2\delta\Lambda_{\rm cc}
+\delta M\,R
-\alpha g_{00}
-\beta R\,g_{00}
-\sigma R^2g_{00}
\right)~.
\label{cc}
\end{equation}
Purely covariant higher-curvature terms, such as $R^2$ and
$R_{\mu\nu}R^{\mu\nu}$, are nonzero on de Sitter, but their
contributions to the metric field equations vanish on an exact
de Sitter solution in four dimensions. They therefore do not
generate independent contributions to the homogeneous background
equations at the order considered.
In this work, however, we use the counterterms written directly
in terms of the lapse and extrinsic curvature and derive their
contributions in ADM variables. The coefficients of the counterterms in the alternative basis~\eqref{cc} must be determined by varying the corresponding action. We leave this calculation for future work.
 
Once the counterterms in Eq. \eqref{eq:Lcc} are included, the equation of motion for $h_{ij}$ and the Hamiltonian constraint must be rederived. 
  The final result is that, with this class of counterterms, the two one-loop Friedmann equations are modified according to the replacements 
\begin{flalign}
    \Lambda_{\rm cc}\rightarrow (\Lambda_{\rm cc})_\text{ph}+\delta \Lambda_{\rm cc}-
    \frac{3\alpha}{2}-3  H^2 (\delta M-\beta)+18 \sigma H^4   \qquad \qquad\text{(First Friedmann equation)}, \\
    \Lambda_{\rm cc}\rightarrow (\Lambda_{\rm cc})_\text{ph}+\delta \Lambda_{\rm cc}-\alpha-3H^2\delta M +36 \sigma H^4  \qquad \qquad \text{(Second Friedmann equation)}~.
\end{flalign}
Divergences can be canceled by choosing the coefficients of the covariant counterterms as
\begin{flalign}
\delta \Lambda_{\rm cc}=\left(\frac{\Lambda^4_{\rm UV}}{16 \pi^2}\right)_\psi+\left(\frac{\Lambda^4_{\rm UV}}{8 \pi^2}\right)_\gamma,
\qquad
\delta M=\left(\frac{\Lambda^2_{\rm UV}}{72\pi^2}\right)_\psi+\left(\frac{\Lambda^2_{\rm UV}}{36\pi^2}\right)_\gamma.
\end{flalign}
The remaining counterterms are fixed by the requirement that the renormalized equations satisfy the Bianchi identities. With the regulator used here, this gives
\begin{equation}
\alpha=\left(\frac{\Lambda^4_{\rm UV}}{12\pi^2}\right)_\psi+\left(\frac{\Lambda^4_{\rm UV}}{6\pi^2}\right)_\gamma,
\quad
\beta=\left(-\frac{\Lambda^2_{\rm UV}}{36\pi^2}\right)_\psi+\left(-\frac{\Lambda^2_{\rm UV}}{18\pi^2}\right)_\gamma,
\quad
\sigma=\left(\frac{1}{96\pi^2}\right)_\psi+\left(\frac{1}{48\pi^2}\right)_\gamma.
\end{equation}
The key result is that, within this non-covariant counterterm structure, the one-loop background equations can be renormalized consistently using background-independent coefficients. In particular, by plugging everything in, we get that the renormalized Hubble parameter reads
\begin{equation}
    \dot{\rho}^2_{\text{1-loop}}=\frac{(\Lambda_{\text{cc}})_{\rm ph}}{3}+\left(\frac{H^4}{8\pi^2M^2_{\rm pl}}\right)_\psi+\left(\frac{7H^4}{12\pi^2M^2_{\rm pl}}\right)_\gamma~,
    \label{eq:renbck}
\end{equation}
where we have temporarily restored $M_{\rm pl}$.
The finite $H^4$ terms are time-independent and have a de Sitter-invariant form. The conclusion is the absence of secular evolution at this order.

The same background-independent renormalization is not possible if the theory is regulated with a fixed comoving cutoff $\Lambda_c$.
Focusing on the scalar part, we find
\begin{flalign}
\dot{\rho}^2=\frac{\Lambda_{\rm cc}}{3}+\frac{a^{-4}\Lambda_c^4+a^{-2}\Lambda_c^2 H^2}{48 \pi ^2}+\ldots~, \\
\ddot{\rho}+3\dot{\rho}^2=\Lambda_{\rm cc}+\frac{a^{-4} \Lambda_c ^4 +2a^{-2}H^2\Lambda_c^2}{48 \pi ^2 }+\ldots~.
\label{eq:1loopcc}
\end{flalign}
where the ellipses denote the corresponding tensor contribution.
Also in this case, the divergences cannot be absorbed by covariant geometric counterterms alone. However, for a fixed comoving cutoff the situation is worse: the required counterterms cannot be chosen with background-independent coefficients. More importantly, their coefficients must be explicitly time-dependent. 

Therefore, in addition to producing loop corrections to the power spectrum that do not freeze on superhorizon scales~\cite{Senatore:2009cf}, a fixed comoving cutoff prevents the divergences in the background equations from being absorbed into counterterms with background-independent, time-independent coefficients. This provides a complementary argument in favor of using a fixed physical cutoff. 

\subsection{Independence of the result from the graviton parametrization}\label{sec:gravNL}

The absence of secular behavior in the Friedmann equations can be traced to the fact that the fluctuation correlators entering these equations always appear with time or spatial derivatives. Undifferentiated correlators such as
\begin{equation}
\langle \gamma_{ij}\gamma^{ij}\rangle
=8\int_{L^{-1}}^{a(t)\Lambda_\text{UV}}\frac{k^2dk}{2\pi^2}|\gamma_{k}|^2=
\frac{\Lambda_{\rm {UV}}^2}{\pi^2}+\frac{2H^2}{\pi^2}\left(\log(\Lambda_{\rm {UV}} L)+Ht\right)~,
\label{secular}
\end{equation}
would induce secular growth if they entered the background equations, as discussed in Sec.~\ref{sec2}.  However, the secular growth is converted into a time-independent contribution when the correlator is acted on by the derivatives that enter the Friedmann equations. 

This result must be independent of the parametrization chosen for the graviton fluctuation. However, as we now discuss, certain parametrizations can obscure this fact by inducing perturbative secular terms in the background variables, which must then be resummed. Consider the exponential parametrization of the graviton
\begin{equation}
{h}_{ij}=e^{2
\tilde{\rho}}\left(e^{\gamma}\right)_{ij}\simeq e^{2\tilde{\rho}}\left(\delta_{ij}+\gamma_{ij}+\frac{1}{2}\gamma_{ik}\gamma_{k  j}+\ldots\right)~,
\label{eq:exppar}
\end{equation}
which is such that $\det [(e^ \gamma)_{ij}]=1$.
In this parametrization, $\dot{\tilde{\rho}}$ cannot be identified with the corresponding quantity in the linear parametrization. By itself, it does not contain all the information about the metric one-point function. In fact, the expectation value of $h_{ij}$ receives additional contributions from the exponential. In particular, restricting to one-loop terms, the quadratic term in Eq.~\eqref{eq:exppar} gives
\begin{equation}
    \langle {h}_{ij}\rangle_\text{one-loop}=e^{2\tilde{\rho}}\left(1+\frac{H^3 t}{3 \pi^2 }+\ldots\right)\delta_{ij}~.
\label{eq:1pExp}
\end{equation}
We stress that, in evaluating the correlator, we neglected the divergent and constant terms. These terms only produce a constant shift of the scale factor, which can be absorbed into its arbitrary normalization.

The apparent secular behavior in the metric one-point function in the exponential parametrization arises from coincident graviton correlators of Eq.~\eqref{secular}. If taken at face value, and restoring $M_\text{pl}$, this term would suggest a secular departure from the background and a breakdown of perturbation theory after a time scale $t\sim M_{\rm pl}^2/H^3$. However, this conclusion is misleading, since no analogous breaking appears in the linear parametrization. To see that this parametrization requires resummation, we derive its associated first Friedmann equation:
\begin{flalign}
&\dot{\tilde{\rho}}^2=\frac{{\Lambda_\text{cc}}}{3}+\frac{1}{6}\langle\dot{\psi}^2+a^{-2}\partial_i\psi\partial^i \psi\rangle+\frac{1}{24}\langle \dot{\gamma}_{ij}\dot{\gamma}^{ij}+a^{-2}\partial_k\gamma_{ij}\partial^k\gamma^{ij}\rangle~.
\end{flalign}
Thus, there is a missing term compared to the result of the linear parametrization~\eqref{eq:1Fr1}. In particular, the relation between the two parameters reads
\begin{equation}
    \dot{\rho}{^2}=\dot{\tilde{\rho}}{^2}+\frac{1}{6} H\partial_t\langle{\gamma}_{ij}\gamma^{ij}\rangle=\dot{\tilde{\rho}}^2+\frac{H^4}{3 \pi^2}~.
\end{equation}
Since the derivative correlators entering the Friedmann equations are time-independent at this order, the corrected expansion rate is constant and can be integrated to reconstruct the metric one-point function. In the two parametrizations, the expectation value of the metric is thus expressed as
\begin{equation}
    \langle
    h_{ij}\rangle=e^{{2{\rho} }}\delta_{ij}=\exp(2{\sqrt{\dot{\tilde{\rho}}^2+\frac{H^4}{3 \pi^2}}}t)\delta_{ij}\sim e^{{2\tilde{\rho} }}\left(1+\frac{H^3}{3\pi^2}t+\ldots\right)\delta_{ij},
\end{equation}
matching Eq.~\eqref{eq:1pExp}.
This shows that part of the secular behavior appearing in nonlinear parametrizations is not a genuine instability of the metric one-point function. Instead, it is generated by the perturbative expansion of the nonlinear map between the original field variable $h_{ij}$ and fluctuations $\gamma_{ij}$, and can be resummed once the corresponding higher-order correlators are included.  

This issue could also propagate if such a parametrization is used to evaluate loop corrections to higher-order correlation functions. For example, loop corrections to the scalar two-point function are obtained from the vertex
\begin{equation}
    h^{ij}\partial_i \psi \partial_j \psi \sim e^{-2\rho(t)}(\delta^{ij}-\gamma^{ij}+\frac{1}{2}\gamma^{ik}\gamma_{k}^j+\ldots)\partial_i \psi \partial_j \psi
\end{equation}
These vertices generate infrared-sensitive loop corrections to correlation functions~\cite{Giddings:2010nc}. The discussion above shows how the corresponding secular terms are reorganized in the metric one-point function. Whether an analogous resummation controls the secular terms appearing in higher-order correlators is a distinct question, beyond the scope of this paper.

\subsection{Caveat on the fixed-physical UV cutoff and equations of motion}
\label{sect:cutofftimed}

A constant physical cutoff corresponds to an explicitly time-dependent upper bound in comoving momentum space.
A common way to impose the prescription is to introduce this cutoff directly on the momentum integral when evaluating a given correlator. This procedure, however, requires some care, since time differentiation does not commute with a projection onto a time-dependent momentum domain. This can be illustrated by considering the example of the time derivative of the regulated coincident two-point function. For a massless scalar in de Sitter, differentiating the regulated correlator as a whole gives \begin{equation}\partial_t\langle\psi^2\rangle_\Lambda=\partial_t\int_{L^{-1}}^{a(t)\Lambda_\text{UV}}\frac{d^3k}{(2\pi)^3}|\psi_k(t)|^2=\frac{H^3}{4\pi^2}.\end{equation}By contrast, differentiating only the integrand and subsequently imposing the cutoff yields\begin{equation}\int_{L^{-1}}^{a(t)\Lambda_\text{UV}}\frac{d^3k}{(2\pi)^3}2{\rm Re}\left[\psi_k^*(t)\dot\psi_k(t)\right]=-\frac{H\Lambda_\text{UV}^2}{4\pi^2}.\end{equation}The discrepancy is accounted for by the boundary term generated by the motion of the cutoff surface:\begin{equation}\partial_t\langle\psi^2\rangle_\Lambda=\frac{4\pi \Lambda_{\rm UV} k^2}{(2\pi)^3}\left(\frac{d}{dt}a(t)\right)|\psi_k(t)|^2|_{k=a(t)\Lambda_\text{UV}}+\int_{L^{-1}}^{a(t)\Lambda_\text{UV}}\frac{d^3k}{(2\pi)^3}\partial_t|\psi_k(t)|^2.\end{equation}
Thus, the relevant non-commutativity is not between Fourier transformation and time differentiation themselves, but between time differentiation and the time-dependent projection defining the regulated field.
In Sec.~\ref{sec:one-loopFr}, we accounted for the associated cutoff-surface contributions by expressing correlators involving time derivatives, such as $\langle\dot\psi^{2}\rangle$, in terms of total time derivatives of undifferentiated correlators, such as $\partial_t\langle\psi^2\rangle$ and $\partial_t^2\langle\psi^2\rangle$. In this way, when the standard Bunch--Davies mode functions are used, the motion of the cutoff surface is consistently incorporated into the background equations.

However, a related subtlety arises if the cutoff is imposed directly at the level of the field operator. Consider the regulated field
\begin{equation}
\psi_\Lambda( x,t)=\int\frac{d^3k}{(2\pi)^3}\theta\big(\Lambda_c(t)-k\big)\left[\psi_k(t)a_{ k}e^{i k\cdot x}+\psi_k^*(t)a_{ k}^\dagger e^{-i k\cdot x}\right],\qquad\Lambda_c(t)\equiv e^{\rho(t)}\Lambda_\text{UV} ,\end{equation}
obtained by applying a time-dependent projector to the unregulated field. Now, this operator does not satisfy the same homogeneous operator equation as the unregulated one. In exact de Sitter space, one instead obtains:
\begin{equation}\theta\big(\Lambda_c-k\big)\left[\ddot\psi_k+3\dot\rho\dot\psi_k+e^{-2\rho}k^2\psi_k\right]+\left[2\dot\rho\Lambda_c\dot\psi_k+4\dot\rho^{2}\Lambda_c\psi_k\right]\delta\big(\Lambda_c-k\big)+\dot\rho^{2}\Lambda_c^2\delta'\big(\Lambda_c-k\big)\psi_k =0.\end{equation}
The first term reproduces the usual Bunch--Davies mode equation in the bulk, whereas the remaining terms are supported on the time-dependent cutoff surface.  Determining whether these terms leave finite contributions to the renormalized background equations requires a treatment in which the cutoff is implemented consistently at the operator level and will be investigated in future work.

\section{Gauge Fixing for Fluctuations and Background Slicing}
\label{sect:gaugef}

In Section~\ref{sec2} we argued that one of the main limitations of standard semiclassical computations is that a gauge fixing imposed on the fluctuations need not be compatible with the underlying classical background slicing. This is an issue if one wants to interpret the backreaction, unless one works directly with gauge-invariant observables~\cite{Garriga:2007zk}. This is why it is conceptually cleaner to impose the gauge condition on the full metric, rather than separately on the background and on the fluctuations. In the previous sections, however, the transverse-traceless gauge was imposed only after splitting the metric into a background and its fluctuations. We therefore track how the gauge condition imposed on the fluctuations translates into a slicing condition for the full metric expectation value.

To understand this issue, we consider the setup of Ref.~\cite{Tsamis:1996qm}. Consider the conformal split
\begin{equation}
g_{\mu\nu}=a^2(\tau)\left(\eta_{\mu\nu}+\gamma_{\mu\nu}\right),
\end{equation}
where we have switched back to a covariant notation. Recall that $a(\tau)$ denotes the classical scale factor, so that the perturbative expansion is organized around the classical metric. To define a gauge condition for fluctuations, one can identify the following gauge-fixing functional 
\begin{equation}
F_\nu[\gamma]=a(\tau)\,\eta^{\rho\mu}\left(\partial_\rho\gamma_{\mu \nu}-\frac{1}{2}\partial_\nu\gamma_{\rho \mu}+2a H \delta_{\rho}^0 \gamma_{\mu \nu} \right)~.
\end{equation}
The important point is that derivatives are taken with respect to the coordinates defined by the classical conformal background.
This gives a legitimate gauge-fixing functional for the perturbative BRST quantization of fluctuations. However, it is not obvious that the resulting backreaction can be interpreted in the original conformal FLRW slicing on which fluctuations are defined.

To make this tension explicit, one might naively expect the metric one-point function computed in this perturbative expansion to resum directly into the standard conformal form of a de Sitter metric, with a constant quantum-corrected expansion parameter $\dot\rho$ and corresponding conformal time $\tau_q$. In other words, the expectation value $\langle \gamma_{\mu \nu}\rangle$ could correct the background in the following way:
\begin{equation}
\langle g_{\mu\nu}\rangle
=\frac{1}{\dot{\rho}^2 \tau^2_q}\eta_{\mu\nu},
\qquad
\langle \gamma_{\mu\nu}\rangle
=[(-H\tau)^{2(1-\frac{\dot{\rho}}{H})}-1]\eta_{\mu \nu}~.
\label{eq:onep}
\end{equation}
 In the second expression, we identified the difference between the classical and quantum metric as the expectation value of the fluctuation, rewriting it in terms of the original classical conformal time using $\tau_q^2=\frac{H^2}{\dot{\rho}^2}\exp(-2(\dot{\rho}- H) t) \tau^2$.
The expectation value of the gauge condition would then require
\begin{equation}
\eta^{\rho \mu}\left(\partial_\rho\langle\gamma_{\mu \nu}\rangle-\frac{1}{2}\partial_\nu\langle\gamma_{\rho \mu}\rangle+2a H \delta_{\rho}^0 \langle\gamma_{\mu \nu} \rangle\right)=0
\label{eq:gaugeF}
\end{equation}
This condition is not satisfied by the ansatz above, unless the classical and quantum expansion parameters coincide. However, this is not the case as follows from the discussion on the expectation value of the Hamiltonian constraint of Sect.~\ref{sec:one-loopFr}. The subtlety is that the metric components in Eq.~\eqref{eq:onep} refer to the original coordinates defined using the classical conformal time. Hence, it is too naive to expect backreaction computed in a gauge defined around the classical background to preserve the original conformal form of the metric, as the corresponding form of the quantum-corrected geometry may only be recovered after an appropriate reparametrization.

The situation would be different if the expansion and gauge condition were defined directly around the metric one-point function, using the corresponding quantum conformal coordinates. The tadpole condition $\langle\gamma_{\mu\nu}\rangle=0$ would then be consistent with the corresponding version of the gauge-fixing functional~\eqref{eq:gaugeF}. Whether this quantum-corrected background admits a solution with constant $\dot\rho$ remains a dynamical question, determined by the constraint and evolution equations.

We now illustrate this mechanism explicitly by working in the transverse-traceless gauge for fluctuations. Consider first the parametrization used in the previous sections
: \begin{equation}
h_{ij}=e^{2\rho(t)}\left(\delta_{ij}+\gamma_{ij}\right)~.
\end{equation}
As we have shown explicitly, imposing the transverse-traceless conditions on $\gamma_{ij}$ is compatible with an FLRW slicing in cosmological time, provided that $\rho(t)$ is determined by the expectation value of the Hamiltonian constraint. By contrast, if the split is organized around the classical solution
\begin{equation}
h_{ij}=a^2(t)\left(\delta_{ij}+\gamma_{ij}\right),
\label{eq:classicaldec}
\end{equation}
then imposing the transverse-traceless gauge on $\gamma_{ij}$ generally shifts the description away from the FLRW slicing in cosmological time, as we show now. 

For a spatially homogeneous and isotropic state, the transverse-traceless conditions always imply
\begin{equation}
\langle \gamma_{ij}\rangle=0,
\end{equation}
irrespective of whether the fluctuation is defined around the metric one-point function or around the classical solution. Thus, taking the expectation value of the two decompositions gives
\begin{equation}
 \langle h_{ij}\rangle_{TT}
=
e^{2\rho(t)}\delta_{ij}
\qquad \neq \qquad \langle h_{ij}\rangle_{TT,0}
=
a^2(t)\delta_{ij}
.
\label{eq:inequality}
\end{equation}
Although the transverse-traceless gauge is imposed on the fluctuation in both cases, it corresponds to two different trace conditions on the original spatial metric. In the first case, the trace is fixed by the quantum-corrected scale factor, while in the second it is fixed by the classical one. This is the sense in which the same gauge condition imposed on the fluctuations corresponds to different gauge choices for the full metric one-point function, depending on the background–fluctuation split.

\subsection{Expanding around the classical solution and the gauge choice}
\label{sect:classical}

We now explain how the constraint equations make the above subtlety explicit. 
Consider the classical  parametrization~\eqref{eq:classicaldec}, with $\gamma_{ij}$ subject to the transverse-traceless gauge as mentioned. One might expect the expectation value of $\gamma_{ij}$ to correct the classical scale factor into the quantum one. As mentioned above, however, the gauge choice gives the condition $\langle \gamma_{ij} \rangle = 0$, which gives Eq.~\eqref{eq:inequality}.
Where did the missing backreaction go? We show it is encoded in the lapse function one-point function. 

In the previous sections, we fixed the one-point function of the spatial metric by solving the expectation value of the Hamiltonian constraint with the conditions $\langle{N}\rangle=1$ and $\langle N^j\rangle= 0$. With this choice, $t$ coincides by construction with the proper time of geodesic observers. This is not the case when the parametrization is performed around the classical solution.

To see this, we expand the expectation value of the Hamiltonian constraint around the classical solution without making assumptions on the background value of $N$. We find
\begin{flalign}
&H^2\left(\langle N^2\rangle-1\right)+\frac{1}{6}\langle\dot{\psi}^2+a^{-2}\partial_i\psi\partial^i \psi\rangle+\frac{1}{24}\langle \dot{\gamma_{ij}}\dot{\gamma}^{ij}+a^{-2}\partial_k\gamma_{ij}\partial^k\gamma^{ij}+4 H\partial_t\left({\gamma}_{ij}\gamma^{ij}\right)\rangle=0
\end{flalign}
where we used the classical relation $3H^2=\Lambda_{\rm cc}$. In deriving this expression, we kept the dependence on the one-point function of the lapse only in the classical terms, since corrections to the lapse-multiplying terms that are already quadratic in the fluctuations would contribute beyond one loop. The result shows that the background gauge choice $\langle N\rangle=1$ is not compatible with imposing the traceless condition on the fluctuation $\gamma_{ij}$ when the latter is defined around the classical solution. This is the ADM analog of the mismatch discussed above in the BRST example.

In order to satisfy the constraint, the expectation value of the lapse must deviate from unity. Consider 
\begin{equation}
N=1+\delta \bar{N}+  n ,
\end{equation}
where $\delta \bar{N}$ denotes the \emph{c-number} deviation of the lapse expectation value from unity, while $  {n}$ is the \emph{operator-valued fluctuation} of the lapse obtained by solving the constraint equations and satisfies $\langle {n}\rangle=0$.  The expectation value of the Hamiltonian constraint is then satisfied provided that
\begin{equation*}
\delta \bar{N}=-\frac{\delta H^2}{2H^2},
\end{equation*}
with
\begin{equation*}
    \delta H^2=\frac{1}{6}\langle\dot{\psi}^2+a^{-2}\partial_i\psi\partial^i \psi\rangle+\frac{1}{24}\langle \dot{\gamma_{ij}}\dot{\gamma}^{ij}+a^{-2}\partial_k\gamma_{ij}\partial^k\gamma^{ij}+4 H\partial_t\left({\gamma}_{ij}\gamma^{ij}\right)\rangle~.
\end{equation*}
Hence, in this parametrization, the backreaction is encoded in the fact that the expectation value of the lapse is shifted away from unity.

However, the two descriptions are related by a change of gauge. Consider a time reparametrization $ t\to \tilde t= t+\xi^0( t)$, leaving the spatial coordinates unchanged, chosen such that
\begin{equation}
N=1+\delta \bar{N}+  n
\qquad \longrightarrow \qquad
\tilde N=1+ n' .
\end{equation}
From here on, we denote by $t$ the time coordinate associated with the expansion around the classical solution and by $\tilde t$ the time coordinate in the gauge where the metric one-point function is written in cosmological time. Moreover, $  n'$ denotes the operator-valued fluctuation of the lapse in the new gauge. In general, this fluctuation need not coincide with the original fluctuation $  n$, since the background and fluctuation parts of the lapse are reshuffled by the coordinate transformation. However, at the perturbative order we are working, the difference between $ n$ and $ n'$ can be neglected. 

At one-loop, the required time reparametrization is determined by\footnote{
The lapse function transforms under a linear time diffeomorphism as
\begin{equation}
\tilde N = N(1-\dot{\xi}^0) .
\end{equation}
Starting from $N=1+\delta \bar{N}+ n$ and requiring $\tilde N=1+ n'$, we find
\begin{equation}
\dot{\xi}^0 = \delta \bar{N}+ n- n' .
\end{equation}
At the perturbative order considered here, the difference between $ n$ and $ n'$ can be neglected. Moreover, in de Sitter, $\delta \bar{N}$ is time independent. Integrating the above equation then gives Eq.~\eqref{xi}.
}
\begin{equation}
\xi^0=-\frac{\delta H^2}{2H^2} t~.
\label{xi}
\end{equation}
With this transformation, the expectation value of the spatial metric is mapped to 
\begin{align}
    \langle {h}_{ij}\rangle_{TT,0}=a^2(t) \delta_{ij}\quad \to\quad  \langle \tilde{h}_{ij}\rangle_{TT}=e^{2\left(H+\frac{\delta H^2}{2H}\right)\tilde{t}}\delta_{ij}=e^{2\rho_{\rm 1-loop}(\tilde t)}\delta_{ij}
\end{align}
After this time reparametrization, we recover the cosmological-time parametrization used in the previous sections. In this new gauge the backreaction is no longer encoded in the expectation value of the lapse, but rather in the spatial scale factor, reproducing the parametrization used in the previous sections.

This shows that imposing the transverse-traceless condition on fluctuations defined around the metric one-point function provides a convenient setup for diagnosing de Sitter breaking of the background, while keeping the relation between the fluctuation gauge and the background slicing under control. With the background gauge choice $\langle N\rangle=1$ and $\langle N^j\rangle=0$ imposed at all orders in the loop expansion, the metric one-point function is written in cosmological time, $\langle h_{ij}\rangle=e^{2\rho(t)}\delta_{ij}$, so that $\dot\rho(t)$ directly defines the quantum-corrected expansion parameter.

\subsection{Matching Correlators Across Background Slicings}

We conclude by checking that correlation functions evaluated by expanding around the classical metric can be matched to those obtained by expanding around the quantum-corrected metric through the same time reparametrization generated by Eq.~\eqref{xi}. 

When the expansion is organized around the classical solution, the graviton two-point function, up to one-loop order, reads schematically
\begin{equation}
    \langle \gamma_{ij}(\vec{x},{t}) \gamma^{ij}(\vec{y},{t})\rangle= \langle \gamma_{ij}(\vec{x},{t}) \gamma^{ij}(\vec{y},{t})\rangle_\text{tree}+\langle \gamma_{ij}(\vec{x},{t}) \gamma^{ij}(\vec{y},{t})\rangle_\text{1-loop}
\end{equation}
Here, we have conveniently split the correlation function into its tree-level contribution, obtained from the quadratic Hamiltonian, and its one-loop correction. Notice, however, that the tree-level part is not the one obtained from the standard Bunch-Davies mode functions given in Eq.~\eqref{bunch}. This is because, since $\langle N\rangle=1+\delta \bar{N}$, the quadratic action for the graviton reads
\begin{equation}
     S_2=\frac{1}{8}\int dt \,d^3x\, e^{3H {t}}\left\{(1+\delta \bar{N})^{-1}\dot{\gamma}_{ij}\dot{\gamma}^{ij}-(1+\delta \bar{N})e^{-2H t}\partial_k\gamma_{ij} \partial^k \gamma^{ij}\right\},
\end{equation}
while the residual operator-valued part $  n$ contributes only to higher-order interactions. With this quadratic action, the tree-level contribution becomes
\begin{equation}
    \langle \gamma_{ij}(\vec{x},{t}) \gamma^{ij}(\vec{y},{t})\rangle_\text{tree}
    ={\frac{2e^{-2 H {t}}}{ \pi^2|\vec{x}-\vec{y}|^2}-\frac{2H^2 (1+\delta \bar{N})^{-2}}{\pi^2} \left(\log \left(\frac{|\vec{x}-\vec{y}|}{L}\right)+\gamma_E -1\right)}
\end{equation}
We notice that, at the order we are working, the prefactor in front of the logarithmic term reads
\begin{equation}
    H^2 (1+\delta \bar{N})^{-2}\simeq \dot\rho_{\rm 1-loop}^{\,2}\,.
\end{equation}
This matches the result obtained by expanding around the one-point function, where all scale factors are defined in terms of the full metric one-point function. However, the time-dependent exponential still depends on the classical Hubble parameter $H$, rather than on the one-loop-corrected background. One might try to compensate for this mismatch through the term $\langle \gamma_{ij}(\vec{x},{t}) \gamma^{ij}(\vec{y},{t})\rangle_\text{1-loop}$. At the order considered here, however, this contribution is insensitive to whether it is evaluated around the classical background or the metric one-point function, since replacing the classical background by its one-loop-corrected counterpart changes an already one-loop quantity only at order $\hbar^2$. It therefore cannot account for the order-$\hbar$ mismatch in the time dependence of the tree-level correlator.

As a result, within this slicing there is no way to replace the exponential dependence on $H$ by the quantum-corrected one appearing in the perturbative expansion around the metric one-point function. The two descriptions match only after applying the coordinate transformation in Eq.~\eqref{xi}. Using this transformation, we find
\begin{equation}
   \langle \gamma_{ij}(\vec{x},\tilde{t}) \gamma^{ij}(\vec{y},\tilde{t})\rangle_\text{tree}={\frac{2e^{-2 (H +\frac{\delta H^2}{2H}) \tilde{t}}}{ \pi^2|\vec{x}-\vec{y}|^2}-\frac{2\dot\rho_{\rm 1-loop}^{\,2}}{\pi^2} \left(\log \left(\frac{|\vec{x}-\vec{y}|}{L}\right)+\gamma_E -1\right)},
\label{eq:gammagamma}
\end{equation}
where the quantity in brackets in the exponent is exactly the one-loop-corrected Hubble parameter. The correlator therefore agrees with the one obtained by expanding directly around the metric one-point function.

Summing up, graviton/scalar-mediated loop corrections to cosmological correlators are of the same perturbative order as the change of the background induced by gravitational backreaction. A consistent treatment should therefore keep track of both effects simultaneously. The analysis above shows, however, that doing so can shift the quantum-corrected background away from the slicing associated with the corresponding classical solution. 
The de Sitter example considered here is simple enough that this mismatch reduces to a constant shift of the Hubble parameter. Nevertheless, Eq.~\eqref{eq:gammagamma} already illustrates the important lesson that the Hubble scale controlling the explicit time dependence of the correlator need not coincide, before the appropriate time reparametrization, with the one controlling the amplitude of its infrared-sensitive part. While this difference is exponentially suppressed at the future boundary for the equal-time two-point function, it could become relevant at higher loop order, where the same one-loop-corrected mode functions now enter time integrals in the Dyson series.
\section{Higher-order correlation functions and two-loop backreaction}
\label{sect:higherL}
The ADM framework developed in this paper can be extended directly to higher-loop computations. At this order, the situation becomes more involved: the correlators entering the quantum Friedmann equations can individually acquire secular contributions. The central question is whether these terms cancel in the complete combination entering the background equations. If they do, the de Sitter isometries remain preserved at two loops; if they do not, the background undergoes a genuine quantum breaking of de Sitter invariance. This is the type of effect investigated in Ref.~\cite{Tsamis:1996qm}. Although we do not resolve this question here, we explain how such infrared-sensitive contributions arise within the present framework.

A useful class of contributions for isolating these effects consists of terms involving both scalar and graviton fluctuations. The leading contributions of this type are obtained by expanding the Hamiltonian constraint to quartic order in the field operators and retaining only terms containing at least one power of the scalar field. This yields
\begin{flalign}
\dot{\rho}_\text{2-loop}^2=H^2+&\frac{1}{6}\langle \dot{\psi}^2+e^{-2\rho}\partial_k\psi\partial^k \psi\rangle_{\text{1-loop}}+\frac{1}{6}e^{-2\rho} \langle\gamma^{im}\gamma^{m j}\partial_i\psi\partial_j\psi- \gamma^{ij}\partial_i\psi\partial_j\psi\rangle_\text{tree-level}\nonumber\\&+\frac{1}{24}\langle \dot{\gamma_{ij}}\dot{\gamma}^{ij}+e^{-2\rho}\partial_k\gamma_{ij}\partial^k{\gamma^{ij}}+4 \dot{\rho}\partial_t\left(\gamma_{ij}\gamma^{ij}\right)\rangle_{\text{1-loop, scalar}}+...\qquad ,
\label{eq:2loopbckg}
\end{flalign}
where, for the graviton quadratic correlators, we only need to account for scalar-loop corrections and not for purely gravitational ones.
All these terms vanish when the scalar field is removed. This dependence is manifest for the cubic and quartic correlators. For the quadratic scalar and graviton correlators, it arises instead through graviton and scalar loop corrections, respectively.

The approximation adopted above involves neglecting correlation functions which are cubic and quartic in ${\gamma}_{ij}$ since the corrections they induce do not scale with the number of scalar fields at this specific loop order. We drop corrections that appear by integrating out the lapse function and the shift vector, since it is possible to verify that the tadpole condition imposed on the auxiliary fields implies that a scalar correlation function cannot receive corrections through a tensor loop generated by an auxiliary field, and vice versa\footnote{In particular, as shown in Appendix~\ref{sec:NandNjeq}, the solution to the constraints provides $N=1+n_\psi+n_\gamma$, where $n_\psi$ and $n_\gamma$ are composite operators, quadratic in the scalar and graviton field, and with vanishing expectation value. When evaluating the expectation value of the Hamiltonian constraint, powers of the lapse function and shift vector reduce at this loop order to $\langle N^2\rangle=1+\langle (n_\psi+n_\gamma)^2\rangle\sim 1+\langle n_\psi^2\rangle+\langle n^2_\gamma\rangle+2\langle n_\gamma\rangle \langle n_\psi\rangle
$. Cross terms vanish due to tadpole conditions at this specific loop order. At higher order, instead, cross terms may be non-vanishing.}.

The way to evaluate correlation functions on the right-hand side of Eq.~\eqref{eq:2loopbckg}, and in general all $n$-point correlators, is to define an interaction picture expansion within the ADM framework of this paper. The field redefinition in Eqs.~\eqref{eq:tadN}--\eqref{eq:tadg}, combined with the renormalized solution to the Friedmann equations~\eqref{eq:renbck},  provides a self-consistent background to define the perturbative expansion, as the standard expansion around the classical solution~\cite{Maldacena:2002vr}. For example, inserting this decomposition into the Einstein-Hilbert action, one obtains the following action:
\begin{flalign}
{S}=
\frac{1}{2}&\int d^4x\,e^{3\rho}\left\{
\frac{1}{4}\dot{\gamma}_{ij}\dot{\gamma}^{ij}-\frac{1}{4}e^{-2\rho}\partial_k \gamma_{ij}\partial^k \gamma^{ij}+\left(1-\frac{1}{4}\gamma_{ij}\gamma^{ij}\right)\left(\dot{\psi}^2-e^{-2\rho}\partial_i\psi\partial^i \psi\right)\right.\nonumber\\&\left.+{e^{-2\rho}}\gamma^{ij}\partial_i\psi \partial_j\psi-\,{e^{-2\rho}}\gamma^{im}\gamma^{mj}\partial_i\psi \partial_j\psi-\frac{1}{2}\left(3\dot{\rho}^2+2\ddot{\rho}-\Lambda_\text{cc}\right)\gamma_{ij}\gamma^{ij}\right\}~,
\label{eq:actionDyn}
\end{flalign}
where we truncated the expansion to operators relevant to the required loop order. For simplicity, we have omitted the non-dynamical fluctuations $n$, $\chi$ and $v^i$. Their constraint solutions are composite operators starting at quadratic order in $\gamma_{ij}$ and $\psi$ and must be substituted back into the action to obtain the complete interaction Hamiltonian at the required loop order.

By introducing interaction picture fields and identifying the free and interacting Hamiltonians, respectively $H_0$ and $H_I$, higher-order $n$-point function can be evaluated to the desired loop order. For example, the cubic correlator of Eq.~\eqref{eq:2loopbckg} is evaluated as in Refs.~\cite{Weinberg:2005vy, Adshead:2008gk}
\begin{equation}
\langle\gamma^{ij}(\vec{x},t)\partial_i\psi(\vec{y},t)\partial_j\psi(\vec{z},t)\rangle_\text{tree-level}=-2 \text{Im}\int_{-\infty_-}^t dt' \langle H_I(t')\gamma^{ij}_I(\vec{x},t)\partial_i\psi_I(\vec{y},t)\partial_j\psi_I(\vec{z},t)\rangle~,
\end{equation}
where we generalized to a spatial-point split. Here, the subscript $I$ denotes interaction-picture fields, and the expectation value on the right-hand side is taken in the free Bunch–Davies state $|0_{\rm BD}\rangle$.

The main improvement from the standard in-in expansion is that the free mode functions, the conformal time and vertices are defined with respect to the quantum-corrected scale factor $e^{\rho(t)}$, rather than the prescribed classical solution $a(t)$. For tree computations, one can set $e^{\rho(t)}\to a(t)$, since the correction to $\rho$ is
of higher order in $\hbar$. This is the same truncation employed in Sect.~\ref{sec:one-loopFr}. Once loop corrections to correlators are included,
however, the corresponding correction to the background must be kept as well.
Therefore, a key advantage of this formalism is that it incorporates quantum corrections to the background from the outset, and defines correlation functions around it. A loop expansion is then applied to disentangle this coupling order by order.

Notice that the same local counterterm action introduced in Eq.~\eqref{eq:Lcc} to renormalise the background must also be included in deriving the interaction Hamiltonian above. If our renormalization procedure is consistent, the corresponding new vertices must also cancel the ultraviolet divergences in the $n$-point correlation functions to one-loop order. Still, additional counterterms that do not contribute to the one-point function, and are therefore unconstrained by the homogeneous background equations, could be required to renormalize $n$-point functions, such as those proportional to $R_{\mu\nu}R^{\mu\nu}$ and $R^2$.

In light of these considerations, Eq.~\eqref{eq:2loopbckg} may exhibit infrared divergences when evaluated term by term. To illustrate this, we focus on the quartic contribution, which is straightforward to compute. Since only its tree-level value is required, it can be evaluated using standard Wick contractions, with $\rho$ approximated by the classical scale factor:
\begin{flalign}
e^{-2\rho}\langle \gamma^{im}\gamma^{mj}\partial_i\psi\partial_j\psi\rangle&=4a^{-2}\int\frac{d^3k d^3q}{(2\pi)^6} q^2\sin^2\theta|\psi_q(\tau)|^2|\gamma_k(\tau)|^2\nonumber\\&=\frac{\left(2 H^2 \Lambda_\text{UV}^2+\Lambda_\text{UV} ^4\right) }{48 \pi ^4}\left(2 H^2 \log \left({\Lambda_\text{UV}  L}\right)+\Lambda_\text{UV} ^2+ 2H^3 t\right)
\end{flalign}
where we exploited the identity \begin{equation}
    \sum_{s}\epsilon_s^{im}(k)\epsilon^{mj}_{s}(k)q_iq_j=2q^2\sin^2\theta_{qk},
\end{equation}

To assess whether the two-loop correction to the metric one-point function contains deviations from a de Sitter form, however, one would need to compute the secular contributions to all the correlation functions entering Eq.~\eqref{eq:2loopbckg} and determine whether they cancel in the complete sum. We leave this analysis for future work.

\section{Conclusions}
In this paper, we investigated the gravitational backreaction of quantum fluctuations on de Sitter geometry using the canonical ADM formulation of General Relativity.
Within this framework, we derived the quantum dynamics of the background directly from the expectation values of the Hamiltonian constraint and spatial metric equation of motion. These quantum Friedmann equations determine the metric one-point function in terms of higher-order correlation functions and constitute the one-point sector of the Schwinger–Dyson hierarchy. 

A central result of our analysis is that the inferred backreaction depends on the background-fluctuation split and on the gauge conditions imposed separately on the background and its fluctuations. This dependence arises because the background and the fluctuations are not independent fields, but two parts of a decomposition of the same underlying metric field. When backreaction is inferred from the gauge-dependent one-point function of the metric, its representation therefore depends on how this decomposition and the corresponding gauge conditions are implemented. In particular, imposing the TT gauge on fluctuations around the classical solution leaves the spatial background metric uncorrected and encodes the backreaction in the background value of the lapse, corresponding to a change of background slicing. By contrast, imposing the same gauge on fluctuations around the metric one-point function is compatible with the cosmological-time FLRW slicing and encodes the backreaction in the quantum-corrected expansion rate. At one loop, the two descriptions are related by the time reparametrization and therefore represent the same quantum-corrected geometry.

Moreover, we applied this new ADM framework to a massless spectator scalar field and the physical graviton helicities, and we evaluated their one-loop contributions to the de Sitter background. After renormalization, these contributions modify the relation between the cosmological constant and the Hubble parameter without generating secular evolution. The renormalized geometry therefore remains de Sitter at this order, in agreement with known results. We also showed that the distribution of quantum corrections between the background and the fluctuation fields depends on the way the metric fluctuations are parametrized. 

Finally, we examined the role of the ultraviolet regulator. A fixed comoving cutoff generates divergences whose subtraction would require counterterms with coefficients depending explicitly on time and on the background evolution. A fixed physical cutoff, by contrast, allows the background equations to be renormalized using time-independent and background-independent coefficients. This is the background-level counterpart of the result obtained in Ref.~\cite{Senatore:2009cf}, where a fixed physical cutoff was shown to be necessary for loop corrections to the power spectrum to freeze on superhorizon scales. Nevertheless, because a hard cutoff breaks diffeomorphism invariance, non-diffeomorphism-invariant restoring counterterms are required. At the level of the homogeneous background equations, the corresponding gravitational consistency conditions reduce to the Bianchi identities, which constrain the counterterms entering the different Friedmann equations. 

Several directions remain open, including a fully operator-level implementation of the physical cutoff, as discussed in Sect.~\ref{sect:cutofftimed}, and the extension of the quantum Friedmann equations to higher-loop order discussed in Sect.~\ref{sect:higherL}.

We conclude by commenting on the gauge dependence of our results. A substantial part of this work has been devoted to clarifying to what extent differences between a priori equivalent perturbative expansions, such as those based on $a(t)$ and $\exp(\rho(t))$, arise from gauge mismatches and can be removed by an appropriate gauge transformation. Given the emergence of these ambiguities, computing gauge-invariant observables would provide a more direct approach to the longstanding question of de Sitter stability. A natural candidate, considered in Ref.~\cite{Garriga:2007zk}, is the expectation value of the renormalized scalar curvature $\langle R\rangle$, suitably smeared with a scalar window function, which can serve as a probe of the local expansion rate in a quasi-de Sitter regime. In more modern language, the role of the window function in making the observable diffeomorphism invariant can be fulfilled by an appropriate gravitational dressing of the metric operators~\cite{Donnelly:2015hta, Donnelly:2016rvo, Giddings:2026fhw}. Although this
would yield a result free from gauge ambiguities once appropriately defined,
constructing gauge-invariant observables in quantum gravity is notoriously
difficult. The dressing becomes increasingly involved at higher orders in $M_{\mathrm{pl}}^{-2}$, and constructing the appropriate observable for a two-loop calculation of the background can therefore be technically demanding.

An advantage of the formalism developed here is that we first fix the gauge directly on the quantum-corrected background and only then compute it. This reverses the usual procedure, in which one fixes the gauge on the classical background before incorporating backreaction, a step that can introduce the ambiguities highlighted in this work. Specifically, the tadpole conditions imposed in Eq.~\eqref{eq:tadpole_condition}
 and \eqref{eq:initialN} select a cosmological time slicing order by order in the loop expansion\footnote{Strictly speaking, one should impose $\langle g_{00}\rangle=-1$ and not $\langle N\rangle=1$, but this distinction does not affect the calculation at the order considered in this paper.}. A complementary investigation, which we have initiated here, therefore asks whether, in a quantum theory on de Sitter spacetime, a consistent choice of gauge conditions from the outset allows the expectation value of the metric to take a spatially flat FLRW form in cosmic time with a constant quantum-corrected expansion parameter.

\subsection*{Acknowledgments}
 We thank Lasha Berezhiani
for many helpful discussions.
This work received support from the French government under the France 2030 investment plan, as part of the Initiative d'Excellence d'Aix-Marseille Universit\'e - A*MIDEX (AMX-19-IET-012). It was also supported by the ``action th\'ematique" Cosmology-Galaxies (ATCG) of the CNRS/INSU PN Astro and by the {\it Agence Nationale de la Recherche} under the grant ANR-24-CE31-6963-01.

\appendix

\section{Solving the Hamiltonian and Momentum Constraints at Quadratic Order}\label{sec:NandNjeq}
In this Appendix, we solve the constraint equations up to the second order in the dynamical operators $\gamma_{ij}$ and $\psi$. 
The solution to the Hamiltonian and momentum constraints was derived in \cite{Seery:2006vu} and \cite{Dimastrogiovanni:2008af} for tensor modes parametrized by the spatial metric $h_{ij}=a^2(t)\left(e^{\gamma}\right)_{ij}$, with $a(t)$ the classical scale factor. In this appendix, we rederive the solutions in the parametrization $h_{ij}=e^{2\rho}\left(\delta_{ij}+\gamma_{ij}\right)$. 
 
We start by expanding the Hamiltonian and momentum constraint up to the second order in the dynamical operators and first order in the non-dynamical ones. We obtain
\begin{flalign}
\biggr(-6\dot{\rho}^2+{12{n}}\dot{\rho}^2&+4\dot{\rho}\partial_k\partial^k\chi\biggr)+\dot{\psi}^2+e^{-2\rho}\partial_i\psi\partial^i \psi+\frac{1}{2}e^{-2\rho}\partial_i\partial_j(\gamma_{im}\gamma_{mj})-\frac{1}{2}e^{-2\rho}\partial_k\partial^k(\gamma_{ij}\gamma^{ij})\nonumber\\&+\frac{1}{4}\dot{\gamma}_{ij}\dot{\gamma}^{ij}+\frac{1}{4}e^{-2\rho}\partial_k\gamma_{ij}\partial^k\gamma^{ij}+{\dot{\rho}}\left(\gamma_{ij}\dot{\gamma}^{ij}+\dot{\gamma}_{ij}\gamma^{ij}\right)+2{\Lambda_\text{cc}}=0 \label{eq:lapse2},
\end{flalign}
and
\begin{flalign}
2\dot{\rho}{\partial_j{n}}-\frac{1}{2}e^{-2\rho}\partial_k\partial^k {v}_j=&\dot{\psi}\partial_j\psi+\frac{1}{2}{\gamma}_{mi}\partial_i\dot{\gamma}_{mj}+\frac{1}{4}\dot{\gamma}_{mi}\partial_j \gamma_{mi}-\frac{1}{2}\partial_j\left({\gamma}_{mi}\dot{\gamma}_{im}\right).
\label{eq:shift2}
\end{flalign}
where $v_i=e^{2\rho}\delta_{ij}v^j$. We solve this set of equations perturbatively in ${\psi}$ and ${\gamma}_{ij}$. Let us start by considering the divergence of the momentum constraint and solve for ${n}$. The solution reads
\begin{equation}
{n}(x,t)=\frac{\partial^{-2}}{2\dot{\rho}}\left(\dot{\psi}\partial^2\psi+\partial^i\dot{\psi}\partial_i\psi+\frac{1}{2}\partial^j\gamma_{mi}\partial_i\dot{\gamma}_{mj}+\frac{1}{4}\partial^j(\dot{\gamma}_{mi}\partial_j \gamma_{mi})\right)-\frac{1}{4\dot{\rho}}\left(\gamma_{mi}\dot{\gamma}_{im}-\langle\gamma_{mi}\dot{\gamma}_{im}\rangle\right),
\label{eq:2n}
\end{equation}
where we fixed the arbitrary function $f(t)$ in such a way that the expectation value of ${n}$ vanishes,  consistently with the tadpole condition $\langle \Omega_\text{BD}|{n}|\Omega_\text{BD}\rangle=0$. 
The second step is to apply the Laplacian on the right and left-hand side of the momentum constraint.  Substituting the solution \eqref{eq:2n} and solving for $v_j$, we have
\begin{flalign}
{v}_j(x,t)=2e^{2\rho}{\partial^{-4}}&\left\{\partial_j\dot{\psi}\partial^2\psi-\partial_l\dot{\psi}\partial_j\partial_l\psi+\partial_j\partial_l\dot{\psi}\partial_l\psi-\partial^2\dot{\psi}\partial_j\psi+\frac{1}{4}\partial_j \dot{\gamma}_{mi}\partial^2 \gamma_{mi}-\frac{1}{4}\partial_l \dot{\gamma}_{mi}\partial_j \partial_l \gamma_{mi}\right.\nonumber\\&\left.+\frac{1}{4}\partial_j \partial_l \dot{\gamma}_{mi}\partial_l \gamma_{mi}-\frac{1}{4}\partial^2 \dot{\gamma}_{mi}\partial_j \gamma_{mi}+\frac{1}{2}\partial_j\left(\partial^l\gamma_{mi}\partial_i\dot{\gamma}_{ml}\right)-\frac{1}{2}\partial^2\left(\gamma_{mi}\partial^i\dot{\gamma}_{mj}\right)\right\}.
\label{eq:2v}
\end{flalign}

The solution for ${\chi}$ is derived by examining the Hamiltonian constraint. We begin by considering the expectation value of the Hamiltonian constraint on $|\Omega_\text{BD}\rangle$ and subtracting it from Eq.~\eqref{eq:lapse2}.  Solving for ${\chi}(x,t)$, we obtain the following expression:
\begin{flalign}
{\chi}(x,t)=-\frac{\partial^{-2}}{4\dot{\rho}}\biggr\{&\dot{\psi}^2+e^{-2\rho}\partial_k\psi\partial^k\psi+\frac{1}{4}\dot{\gamma}_{ij}\dot{\gamma}^{ij}+\frac{1}{4}e^{-2\rho}\partial_k\gamma_{ij}\partial^k\gamma^{ij}+{\dot{\rho}}\left(\gamma_{ij}\dot{\gamma}^{ij}+\dot{\gamma}_{ij}\gamma^{ij}\right)\nonumber\\&+\frac{1}{2}e^{-2\rho}\partial_i\partial_j(\gamma_{im}\gamma_{mj})-\frac{1}{2}e^{-2\rho}\partial_k\partial^k(\gamma_{ij}\gamma^{ij})+12\dot{\rho}^2{n}-\langle \dot{\psi}^2+e^{-2\rho}\partial_k\psi\partial^k\psi\nonumber\\&+\frac{1}{4}\dot\gamma_{ij}\dot\gamma^{ij}+\frac{1}{4}e^{-2\rho}\partial_k\gamma_{ij}\partial^k\gamma^{ij}+{\dot{\rho}}\left(\gamma_{ij}\dot{\gamma}^{ij}+\dot{\gamma}_{ij}\gamma^{ij}\right)\rangle\biggr\},
\label{eq:2chi}
\end{flalign}
with ${n}$ determined by \eqref{eq:2n}. With this, we have found how to write the fluctuations of the lapse function and shift vector in terms of the dynamical degrees of freedom. These are quadratic in these variables, as expected.

\bibliographystyle{JHEP}
 \bibliography{bibl.bib}

@article{Weinberg:2005vy,
    author = "Weinberg, Steven",
    title = "{Quantum contributions to cosmological correlations}",
    eprint = "hep-th/0506236",
    archivePrefix = "arXiv",
    reportNumber = "UTTG-01-05",
    doi = "10.1103/PhysRevD.72.043514",
    journal = "Phys. Rev. D",
    volume = "72",
    pages = "043514",
    year = "2005"
}

@article{Boyanovsky:2005sh,
    author = "Boyanovsky, D. and de Vega, Hector J. and Sanchez, Norma G.",
    title = "{Quantum corrections to slow roll inflation and new scaling of superhorizon fluctuations}",
    eprint = "astro-ph/0503669",
    archivePrefix = "arXiv",
    doi = "10.1016/j.nuclphysb.2006.04.010",
    journal = "Nucl. Phys. B",
    volume = "747",
    pages = "25--54",
    year = "2006"
}

@article{Boyanovsky:2004ph,
    author = "Boyanovsky, D. and de Vega, Hector J. and Sanchez, Norma G.",
    title = "{Particle decay during inflation: Self-decay of inflaton quantum fluctuations during slow roll}",
    eprint = "astro-ph/0409406",
    archivePrefix = "arXiv",
    doi = "10.1103/PhysRevD.71.023509",
    journal = "Phys. Rev. D",
    volume = "71",
    pages = "023509",
    year = "2005"
}

@article{Antoniadis:1986sb,
    author = "Antoniadis, Ignatios and Mottola, E.",
    title = "{Graviton Fluctuations in De Sitter Space}",
    reportNumber = "CERN-TH-4605/86",
    doi = "10.1063/1.529381",
    journal = "J. Math. Phys.",
    volume = "32",
    pages = "1037--1044",
    year = "1991"
}

@article{Miao:2017vly,
    author = "Miao, S. P. and Tsamis, N. C. and Woodard, R. P.",
    title = "{Invariant measure of the one-loop quantum gravitational backreaction on inflation}",
    eprint = "1702.05694",
    archivePrefix = "arXiv",
    primaryClass = "gr-qc",
    reportNumber = "UFIFT-QG-17-01, CCTP-2017-1",
    doi = "10.1103/PhysRevD.95.125008",
    journal = "Phys. Rev. D",
    volume = "95",
    number = "12",
    pages = "125008",
    year = "2017"
}

@article{Pimentel:2012tw,
    author = "Pimentel, Guilherme L. and Senatore, Leonardo and Zaldarriaga, Matias",
    title = "{On Loops in Inflation III: Time Independence of zeta in Single Clock Inflation}",
    eprint = "1203.6651",
    archivePrefix = "arXiv",
    primaryClass = "hep-th",
    doi = "10.1007/JHEP07(2012)166",
    journal = "JHEP",
    volume = "07",
    pages = "166",
    year = "2012"
}

@article{Braglia:2025qrb,
    author = "Braglia, Matteo and Pinol, Lucas",
    title = "{Freezing of the renormalized one-loop primordial scalar power spectrum}",
    eprint = "2504.13136",
    archivePrefix = "arXiv",
    primaryClass = "astro-ph.CO",
    doi = "10.1103/5jkv-mj3w",
    journal = "Phys. Rev. D",
    volume = "113",
    number = "6",
    pages = "L061302",
    year = "2026"
}

@article{Allen:1985ux,
    author = "Allen, Bruce",
    title = "{Vacuum States in de Sitter Space}",
    reportNumber = "UCSB-TH-3-1985",
    doi = "10.1103/PhysRevD.32.3136",
    journal = "Phys. Rev. D",
    volume = "32",
    pages = "3136",
    year = "1985"
}

@article{Allen:1986ta,
    author = "Allen, Bruce",
    title = "{The Graviton Propagator in De Sitter Space}",
    reportNumber = "TUTP-86-9",
    doi = "10.1103/PhysRevD.34.3670",
    journal = "Phys. Rev. D",
    volume = "34",
    pages = "3670",
    year = "1986"
}

@article{Higuchi:2011vw,
    author = "Higuchi, Atsushi and Marolf, Donald and Morrison, Ian A.",
    title = "{de Sitter invariance of the dS graviton vacuum}",
    eprint = "1107.2712",
    archivePrefix = "arXiv",
    primaryClass = "hep-th",
    doi = "10.1088/0264-9381/28/24/245012",
    journal = "Class. Quant. Grav.",
    volume = "28",
    pages = "245012",
    year = "2011"
}

@article{Polyakov:2007mm,
    author = "Polyakov, A. M.",
    title = "{De Sitter space and eternity}",
    eprint = "0709.2899",
    archivePrefix = "arXiv",
    primaryClass = "hep-th",
    reportNumber = "PUPT-2244",
    doi = "10.1016/j.nuclphysb.2008.01.002",
    journal = "Nucl. Phys. B",
    volume = "797",
    pages = "199--217",
    year = "2008"
}

@article{Polyakov:2009nq,
    author = "Polyakov, A. M.",
    title = "{Decay of Vacuum Energy}",
    eprint = "0912.5503",
    archivePrefix = "arXiv",
    primaryClass = "hep-th",
    reportNumber = "PUPT-2320",
    doi = "10.1016/j.nuclphysb.2010.03.021",
    journal = "Nucl. Phys. B",
    volume = "834",
    pages = "316--329",
    year = "2010"
}

@article{Polyakov:2012uc,
    author = "Polyakov, A. M.",
    title = "{Infrared instability of the de Sitter space}",
    eprint = "1209.4135",
    archivePrefix = "arXiv",
    primaryClass = "hep-th",
    month = "9",
    year = "2012"
}

@article{Dvali:2017eba,
    author = "Dvali, Gia and Gomez, Cesar and Zell, Sebastian",
    title = "{Quantum Break-Time of de Sitter}",
    eprint = "1701.08776",
    archivePrefix = "arXiv",
    primaryClass = "hep-th",
    reportNumber = "LMU-ASC-08-17, MPP-2017-10, LMU-ASC 08/17, MPP-2017-10",
    doi = "10.1088/1475-7516/2017/06/028",
    journal = "JCAP",
    volume = "06",
    pages = "028",
    year = "2017"
}

@article{Berezhiani:2016grw,
    author = "Berezhiani, Lasha",
    title = "{On Corpuscular Theory of Inflation}",
    eprint = "1610.08433",
    archivePrefix = "arXiv",
    primaryClass = "hep-th",
    doi = "10.1140/epjc/s10052-017-4672-5",
    journal = "Eur. Phys. J. C",
    volume = "77",
    number = "2",
    pages = "106",
    year = "2017"
}

@article{Giddings:2010nc,
    author = "Giddings, Steven B. and Sloth, Martin S.",
    title = "{Semiclassical relations and IR effects in de Sitter and slow-roll space-times}",
    eprint = "1005.1056",
    archivePrefix = "arXiv",
    primaryClass = "hep-th",
    reportNumber = "CERN-PH-TH-2010-095",
    doi = "10.1088/1475-7516/2011/01/023",
    journal = "JCAP",
    volume = "01",
    pages = "023",
    year = "2011"
}

@article{Senatore:2009cf,
    author = "Senatore, Leonardo and Zaldarriaga, Matias",
    title = "{On Loops in Inflation}",
    eprint = "0912.2734",
    archivePrefix = "arXiv",
    primaryClass = "hep-th",
    doi = "10.1007/JHEP12(2010)008",
    journal = "JHEP",
    volume = "12",
    pages = "008",
    year = "2010"
}

@article{Tsamis:1996qq,
    author = "Tsamis, N. C. and Woodard, R. P.",
    title = "{Quantum gravity slows inflation}",
    eprint = "hep-ph/9602315",
    archivePrefix = "arXiv",
    reportNumber = "CPTH-S372-0995, CRETE-95-11, UFIFT-HEP-95-17",
    doi = "10.1016/0550-3213(96)00246-5",
    journal = "Nucl. Phys. B",
    volume = "474",
    pages = "235--248",
    year = "1996"
}

@article{Garriga:2007zk,
    author = "Garriga, Jaume and Tanaka, Takahiro",
    title = "{Can infrared gravitons screen Lambda?}",
    eprint = "0706.0295",
    archivePrefix = "arXiv",
    primaryClass = "hep-th",
    reportNumber = "KUNS-2076",
    doi = "10.1103/PhysRevD.77.024021",
    journal = "Phys. Rev. D",
    volume = "77",
    pages = "024021",
    year = "2008"
}

@article{Tsamis:1996qm,
    author = "Tsamis, N. C. and Woodard, R. P.",
    title = "{The Quantum gravitational back reaction on inflation}",
    eprint = "hep-ph/9602316",
    archivePrefix = "arXiv",
    reportNumber = "CPTH-S422-1295, CRETE-96-13, UFIFT-HEP-96-5",
    doi = "10.1006/aphy.1997.5613",
    journal = "Annals Phys.",
    volume = "253",
    pages = "1--54",
    year = "1997"
}

@article{Ford:1984hs,
    author = "Ford, L. H.",
    title = "{Quantum Instability of De Sitter Space-time}",
    reportNumber = "IMPERIAL-TP-83-84-52",
    doi = "10.1103/PhysRevD.31.710",
    journal = "Phys. Rev. D",
    volume = "31",
    pages = "710",
    year = "1985"
}

@article{Starobinsky:1994bd,
    author = "Starobinsky, Alexei A. and Yokoyama, Junichi",
    title = "{Equilibrium state of a selfinteracting scalar field in the De Sitter background}",
    eprint = "astro-ph/9407016",
    archivePrefix = "arXiv",
    reportNumber = "YITP-U-94-12",
    doi = "10.1103/PhysRevD.50.6357",
    journal = "Phys. Rev. D",
    volume = "50",
    pages = "6357--6368",
    year = "1994"
}

@article{Starobinsky:1986fx,
    author = "Starobinsky, Alexei A.",
    title = "{STOCHASTIC DE SITTER (INFLATIONARY) STAGE IN THE EARLY UNIVERSE}",
    doi = "10.1007/3-540-16452-9_6",
    journal = "Lect. Notes Phys.",
    volume = "246",
    pages = "107--126",
    year = "1986"
}

@article{Gorbenko:2019rza,
    author = "Gorbenko, Victor and Senatore, Leonardo",
    title = "{$\lambda \phi^4$ in dS}",
    eprint = "1911.00022",
    archivePrefix = "arXiv",
    primaryClass = "hep-th",
    month = "10",
    year = "2019"
}

@article{Baumgart:2019clc,
    author = "Baumgart, Matthew and Sundrum, Raman",
    title = "{De Sitter Diagrammar and the Resummation of Time}",
    eprint = "1912.09502",
    archivePrefix = "arXiv",
    primaryClass = "hep-th",
    doi = "10.1007/JHEP07(2020)119",
    journal = "JHEP",
    volume = "07",
    pages = "119",
    year = "2020"
}

@article{Berezhiani:2024boz,
    author = "Berezhiani, Lasha and Dvali, Gia and Sakhelashvili, Otari",
    title = "{Consistent canonical quantization of gravity: Recovery of classical GR from BRST-invariant coherent states}",
    eprint = "2409.18777",
    archivePrefix = "arXiv",
    primaryClass = "hep-th",
    doi = "10.1103/652w-ym62",
    journal = "Phys. Rev. D",
    volume = "113",
    number = "12",
    pages = "125032",
    year = "2026"
}

@article{Berezhiani:2021gph,
    author = "Berezhiani, Lasha and Cintia, Giordano and Zantedeschi, Michael",
    title = "{Background-field method and initial-time singularity for coherent states}",
    eprint = "2108.13235",
    archivePrefix = "arXiv",
    primaryClass = "hep-th",
    doi = "10.1103/PhysRevD.105.045003",
    journal = "Phys. Rev. D",
    volume = "105",
    number = "4",
    pages = "045003",
    year = "2022"
}

@article{Berezhiani:2023uwt,
    author = "Berezhiani, Lasha and Cintia, Giordano and Zantedeschi, Michael",
    title = "{Perturbative construction of coherent states}",
    eprint = "2311.18650",
    archivePrefix = "arXiv",
    primaryClass = "hep-th",
    doi = "10.1103/PhysRevD.109.085018",
    journal = "Phys. Rev. D",
    volume = "109",
    number = "8",
    pages = "085018",
    year = "2024"
}

@article{Dvali:2013vxa,
    author = "Dvali, Gia and Flassig, Daniel and Gomez, Cesar and Pritzel, Alexander and Wintergerst, Nico",
    title = "{Scrambling in the Black Hole Portrait}",
    eprint = "1307.3458",
    archivePrefix = "arXiv",
    primaryClass = "hep-th",
    reportNumber = "LMU-ASC-50-13, LMU-ASC 50/13",
    doi = "10.1103/PhysRevD.88.124041",
    journal = "Phys. Rev. D",
    volume = "88",
    number = "12",
    pages = "124041",
    year = "2013"
}

@article{Dvali:2012en,
    author = "Dvali, Gia and Gomez, Cesar",
    title = "{Black Holes as Critical Point of Quantum Phase Transition}",
    eprint = "1207.4059",
    archivePrefix = "arXiv",
    primaryClass = "hep-th",
    doi = "10.1140/epjc/s10052-014-2752-3",
    journal = "Eur. Phys. J. C",
    volume = "74",
    pages = "2752",
    year = "2014"
}

@article{Dvali:2013eja,
    author = "Dvali, Gia and Gomez, Cesar",
    title = "{Quantum Compositeness of Gravity: Black Holes, AdS and Inflation}",
    eprint = "1312.4795",
    archivePrefix = "arXiv",
    primaryClass = "hep-th",
    doi = "10.1088/1475-7516/2014/01/023",
    journal = "JCAP",
    volume = "01",
    pages = "023",
    year = "2014"
}

@article{Berezhiani:2021zst,
    author = "Berezhiani, Lasha and Dvali, Gia and Sakhelashvili, Otari",
    title = "{de Sitter space as a BRST invariant coherent state of gravitons}",
    eprint = "2111.12022",
    archivePrefix = "arXiv",
    primaryClass = "hep-th",
    doi = "10.1103/PhysRevD.105.025022",
    journal = "Phys. Rev. D",
    volume = "105",
    number = "2",
    pages = "025022",
    year = "2022"
}

@article{Berezhiani:2024pub,
    author = "Berezhiani, Lasha and Dvali, Gia and Sakhelashvili, Otari",
    title = "{Coherent states in gauge theories: Topological defects and other classical configurations}",
    eprint = "2411.11657",
    archivePrefix = "arXiv",
    primaryClass = "hep-th",
    doi = "10.1103/PhysRevD.111.065018",
    journal = "Phys. Rev. D",
    volume = "111",
    number = "6",
    pages = "065018",
    year = "2025"
}

@article{Dvali:2017ruz,
    author = "Dvali, Gia and Zell, Sebastian",
    title = "{Classicality and Quantum Break-Time for Cosmic Axions}",
    eprint = "1710.00835",
    archivePrefix = "arXiv",
    primaryClass = "hep-ph",
    reportNumber = "LMU-ASC-59-17, LMU-ASC 59/17",
    doi = "10.1088/1475-7516/2018/07/064",
    journal = "JCAP",
    volume = "07",
    pages = "064",
    year = "2018"
}

@article{Dvali:2022vzz,
    author = "Dvali, Gia and Eisemann, Lukas",
    title = "{Perturbative understanding of nonperturbative processes and quantumization versus classicalization}",
    eprint = "2211.02618",
    archivePrefix = "arXiv",
    primaryClass = "hep-th",
    doi = "10.1103/PhysRevD.106.125019",
    journal = "Phys. Rev. D",
    volume = "106",
    number = "12",
    pages = "125019",
    year = "2022"
}

@article{Berezhiani:2022gnv,
    author = "Berezhiani, Lasha and Trodden, Mark",
    title = "{A relativistic gas of inflatons as an initial state for inflation}",
    eprint = "2211.06222",
    archivePrefix = "arXiv",
    primaryClass = "hep-th",
    doi = "10.1016/j.physletb.2023.137852",
    journal = "Phys. Lett. B",
    volume = "840",
    pages = "137852",
    year = "2023"
}

@article{Berezhiani:2015ola,
    author = "Berezhiani, Lasha and Trodden, Mark",
    title = "{How Likely are Constituent Quanta to Initiate Inflation?}",
    eprint = "1504.01730",
    archivePrefix = "arXiv",
    primaryClass = "hep-th",
    doi = "10.1016/j.physletb.2015.08.007",
    journal = "Phys. Lett. B",
    volume = "749",
    pages = "425--430",
    year = "2015"
}

@article{Berezhiani:2020pbv,
    author = "Berezhiani, Lasha and Zantedeschi, Michael",
    title = "{Evolution of coherent states as quantum counterpart of classical dynamics}",
    eprint = "2011.11229",
    archivePrefix = "arXiv",
    primaryClass = "hep-th",
    doi = "10.1103/PhysRevD.104.085007",
    journal = "Phys. Rev. D",
    volume = "104",
    number = "8",
    pages = "085007",
    year = "2021"
}

@article{Franciolini:2023agm,
    author = "Franciolini, Gabriele and Iovino, Junior., Antonio and Taoso, Marco and Urbano, Alfredo",
    title = "{Perturbativity in the presence of ultraslow-roll dynamics}",
    eprint = "2305.03491",
    archivePrefix = "arXiv",
    primaryClass = "astro-ph.CO",
    doi = "10.1103/PhysRevD.109.123550",
    journal = "Phys. Rev. D",
    volume = "109",
    number = "12",
    pages = "123550",
    year = "2024"
}

@article{Kristiano:2022maq,
    author = "Kristiano, Jason and Yokoyama, Jun'ichi",
    title = "{Constraining Primordial Black Hole Formation from Single-Field Inflation}",
    eprint = "2211.03395",
    archivePrefix = "arXiv",
    primaryClass = "hep-th",
    reportNumber = "RESCEU-20/22",
    doi = "10.1103/PhysRevLett.132.221003",
    journal = "Phys. Rev. Lett.",
    volume = "132",
    number = "22",
    pages = "221003",
    year = "2024"
}

@article{Inomata:2025bqw,
    author = "Inomata, Keisuke",
    title = "{Conservation of superhorizon curvature perturbations at one loop: Backreaction in the in-in formalism and renormalization}",
    eprint = "2502.08707",
    archivePrefix = "arXiv",
    primaryClass = "astro-ph.CO",
    doi = "10.1103/PhysRevD.111.103504",
    journal = "Phys. Rev. D",
    volume = "111",
    number = "10",
    pages = "103504",
    year = "2025"
}

@article{Boyanovsky:2005px,
    author = "Boyanovsky, D. and de Vega, Hector J. and Sanchez, Norma G.",
    title = "{Quantum corrections to the inflaton potential and the power spectra from superhorizon modes and trace anomalies}",
    eprint = "astro-ph/0507596",
    archivePrefix = "arXiv",
    doi = "10.1103/PhysRevD.72.103006",
    journal = "Phys. Rev. D",
    volume = "72",
    pages = "103006",
    year = "2005"
}

@article{Abramo:1997hu,
    author = "Abramo, L. Raul W. and Brandenberger, Robert H. and Mukhanov, Viatcheslav F.",
    title = "{The Energy - momentum tensor for cosmological perturbations}",
    eprint = "gr-qc/9704037",
    archivePrefix = "arXiv",
    reportNumber = "BROWN-HET-1046",
    doi = "10.1103/PhysRevD.56.3248",
    journal = "Phys. Rev. D",
    volume = "56",
    pages = "3248--3257",
    year = "1997"
}

@article{Mukhanov:1996ak,
    author = "Mukhanov, Viatcheslav F. and Abramo, L. Raul W. and Brandenberger, Robert H.",
    title = "{On the Back reaction problem for gravitational perturbations}",
    eprint = "gr-qc/9609026",
    archivePrefix = "arXiv",
    reportNumber = "BROWN-HET-1045",
    doi = "10.1103/PhysRevLett.78.1624",
    journal = "Phys. Rev. Lett.",
    volume = "78",
    pages = "1624--1627",
    year = "1997"
}

@article{Maldacena:2002vr,
    author = "Maldacena, Juan Martin",
    title = "{Non-Gaussian features of primordial fluctuations in single field inflationary models}",
    eprint = "astro-ph/0210603",
    archivePrefix = "arXiv",
    doi = "10.1088/1126-6708/2003/05/013",
    journal = "JHEP",
    volume = "05",
    pages = "013",
    year = "2003"
}

@article{Berezhiani:2025roh,
    author = "Berezhiani, Lasha and Dvali, Gia and Sakhelashvili, Otari",
    title = "{On Time-Evolution in Quantum Gravity}",
    eprint = "2510.16543",
    archivePrefix = "arXiv",
    primaryClass = "hep-th",
    month = "10",
    year = "2025"
}

@book{Birrell:1982ix,
    author = "Birrell, N. D. and Davies, P. C. W.",
    title = "{Quantum Fields in Curved Space}",
    doi = "10.1017/CBO9780511622632",
    isbn = "978-0-511-62263-2, 978-0-521-27858-4",
    publisher = "Cambridge University Press",
    address = "Cambridge, UK",
    series = "Cambridge Monographs on Mathematical Physics",
    year = "1982"
}

@article{Berezhiani:2025tkp,
    author = "Berezhiani, Lasha and Cintia, Giordano and Contri, Giacomo",
    title = "{Coherence and Quantum Stability of Relativistic Superfluid States}",
    eprint = "2509.21667",
    archivePrefix = "arXiv",
    primaryClass = "hep-th",
    month = "9",
    year = "2025"
}

@article{Bunch:1978yq,
    author = "Bunch, T. S. and Davies, P. C. W.",
    title = "{Quantum Field Theory in de Sitter Space: Renormalization by Point Splitting}",
    doi = "10.1098/rspa.1978.0060",
    journal = "Proc. Roy. Soc. Lond. A",
    volume = "360",
    pages = "117--134",
    year = "1978"
}

@article{Seery:2006vu,
    author = "Seery, David and Lidsey, James E. and Sloth, Martin S.",
    title = "{The inflationary trispectrum}",
    eprint = "astro-ph/0610210",
    archivePrefix = "arXiv",
    doi = "10.1088/1475-7516/2007/01/027",
    journal = "JCAP",
    volume = "01",
    pages = "027",
    year = "2007"
}

@article{Dimastrogiovanni:2008af,
    author = "Dimastrogiovanni, Emanuela and Bartolo, Nicola",
    title = "{One-loop graviton corrections to the curvature perturbation from inflation}",
    eprint = "0807.2790",
    archivePrefix = "arXiv",
    primaryClass = "astro-ph",
    doi = "10.1088/1475-7516/2008/11/016",
    journal = "JCAP",
    volume = "11",
    pages = "016",
    year = "2008"
}

@article{Boyanovsky:1994me,
    author = "Boyanovsky, D. and de Vega, H. J. and Holman, R. and Lee, D. S. and Singh, A.",
    title = "{Dissipation via particle production in scalar field theories}",
    eprint = "hep-ph/9408214",
    archivePrefix = "arXiv",
    reportNumber = "PAR-LPTHE-94-31, LPTHE-94-31, PITT-94-07A, CMU-HEP-94-23, DOE-ER-40682-77",
    doi = "10.1103/PhysRevD.51.4419",
    journal = "Phys. Rev. D",
    volume = "51",
    pages = "4419--4444",
    year = "1995"
}

@article{Cheung:2007st,
    author = "Cheung, Clifford and Creminelli, Paolo and Fitzpatrick, A. Liam and Kaplan, Jared and Senatore, Leonardo",
    title = "{The Effective Field Theory of Inflation}",
    eprint = "0709.0293",
    archivePrefix = "arXiv",
    primaryClass = "hep-th",
    reportNumber = "IC-2007-032",
    doi = "10.1088/1126-6708/2008/03/014",
    journal = "JHEP",
    volume = "03",
    pages = "014",
    year = "2008"
}

@book{Peskin:1995ev,
    author = "Peskin, Michael E. and Schroeder, Daniel V.",
    title = "{An Introduction to quantum field theory}",
    doi = "10.1201/9780429503559",
    isbn = "978-0-201-50397-5, 978-0-429-50355-9, 978-0-429-49417-8",
    publisher = "Addison-Wesley",
    address = "Reading, USA",
    year = "1995"
}

@article{Glauber:1963tx,
    author = "Glauber, Roy J.",
    title = "{Coherent and incoherent states of the radiation field}",
    doi = "10.1103/PhysRev.131.2766",
    journal = "Phys. Rev.",
    volume = "131",
    pages = "2766--2788",
    year = "1963"
}

@article{Kibble:1965zza,
    author = "Kibble, T. W. B.",
    title = "{Frequency Shift in High-Intensity Compton Scattering}",
    doi = "10.1103/PhysRev.138.B740",
    journal = "Phys. Rev.",
    volume = "138",
    pages = "B740--B753",
    year = "1965"
}

@article{Zhang:1990fy,
    author = "Zhang, Wei-Min and Feng, Da Hsuan and Gilmore, Robert",
    title = "{Coherent States: Theory and Some Applications}",
    doi = "10.1103/RevModPhys.62.867",
    journal = "Rev. Mod. Phys.",
    volume = "62",
    pages = "867--927",
    year = "1990"
}

@article{Zhang:1999is,
    author = "Zhang, Wei-Min",
    editor = "Mitra, Asoke N.",
    title = "{Coherent states in field theory}",
    eprint = "hep-th/9908117",
    archivePrefix = "arXiv",
    pages = "297--323",
    month = "8",
    year = "1999"
}

@article{Dvali:2024dzf,
    author = "Dvali, Gia",
    title = "{A string theoretic derivation of gibbons-hawking entropy}",
    eprint = "2407.01510",
    archivePrefix = "arXiv",
    primaryClass = "hep-th",
    doi = "10.1007/s10714-025-03446-6",
    journal = "Gen. Rel. Grav.",
    volume = "57",
    number = "8",
    pages = "118",
    year = "2025"
}

@article{Weinberg:2006ac,
    author = "Weinberg, Steven",
    title = "{Quantum contributions to cosmological correlations. II. Can these corrections become large?}",
    eprint = "hep-th/0605244",
    archivePrefix = "arXiv",
    reportNumber = "UTTG-0306",
    doi = "10.1103/PhysRevD.74.023508",
    journal = "Phys. Rev. D",
    volume = "74",
    pages = "023508",
    year = "2006"
}

@article{Green:2020whw,
    author = "Green, Daniel and Porto, Rafael A.",
    title = "{Signals of a Quantum Universe}",
    eprint = "2001.09149",
    archivePrefix = "arXiv",
    primaryClass = "hep-th",
    doi = "10.1103/PhysRevLett.124.251302",
    journal = "Phys. Rev. Lett.",
    volume = "124",
    number = "25",
    pages = "251302",
    year = "2020"
}

@article{Arnowitt:1962hi,
    author = "Arnowitt, Richard L. and Deser, Stanley and Misner, Charles W.",
    title = "{The Dynamics of general relativity}",
    eprint = "gr-qc/0405109",
    archivePrefix = "arXiv",
    doi = "10.1007/s10714-008-0661-1",
    journal = "Gen. Rel. Grav.",
    volume = "40",
    pages = "1997--2027",
    year = "2008"
}

@article{Piazza:2013coa,
    author = "Piazza, Federico and Vernizzi, Filippo",
    title = "{Effective Field Theory of Cosmological Perturbations}",
    eprint = "1307.4350",
    archivePrefix = "arXiv",
    primaryClass = "hep-th",
    doi = "10.1088/0264-9381/30/21/214007",
    journal = "Class. Quant. Grav.",
    volume = "30",
    pages = "214007",
    year = "2013"
}

@article{Braglia:2025cee,
    author = "Braglia, Matteo and Pinol, Lucas",
    title = "{One-loop renormalization of the effective field theory of inflationary fluctuations from gravitational interactions}",
    eprint = "2504.07926",
    archivePrefix = "arXiv",
    primaryClass = "astro-ph.CO",
    doi = "10.1103/f6q6-5jxb",
    journal = "Phys. Rev. D",
    volume = "113",
    number = "6",
    pages = "063513",
    year = "2026"
}

@article{Fumagalli:2023zzl,
    author = "Fumagalli, Jacopo",
    title = "{Absence of one-loop effects on large scales from small scales in non-slow-roll dynamics}",
    eprint = "2305.19263",
    archivePrefix = "arXiv",
    primaryClass = "astro-ph.CO",
    doi = "10.1007/JHEP05(2025)162",
    journal = "JHEP",
    volume = "05",
    pages = "162",
    year = "2025"
}

@article{Fumagalli:2024jzz,
    author = "Fumagalli, Jacopo",
    title = "{Absence of one-loop effects on large scales from small scales in non-slow-roll dynamics. Part 2. Quartic interactions and consistency relations}",
    eprint = "2408.08296",
    archivePrefix = "arXiv",
    primaryClass = "astro-ph.CO",
    doi = "10.1007/JHEP01(2025)108",
    journal = "JHEP",
    volume = "01",
    pages = "108",
    year = "2025"
}

@article{Janssen:2008dw,
    author = "Janssen, Tomas and Prokopec, Tomislav",
    title = "{The Graviton one-loop effective action in cosmological space-times with constant deceleration}",
    eprint = "0807.0447",
    archivePrefix = "arXiv",
    primaryClass = "gr-qc",
    reportNumber = "ITP-UU-08-38, SPIN-08-29",
    doi = "10.1016/j.aop.2009.09.003",
    journal = "Annals Phys.",
    volume = "325",
    pages = "948--968",
    year = "2010"
}

@article{Caravano:2024moy,
    author = "Caravano, Angelo and Franciolini, Gabriele and Renaux-Petel, S{\'e}bastien",
    title = "{Ultraslow-roll inflation on the lattice: Backreaction and nonlinear effects}",
    eprint = "2410.23942",
    archivePrefix = "arXiv",
    primaryClass = "astro-ph.CO",
    reportNumber = "CERN-TH-2024-181",
    doi = "10.1103/PhysRevD.111.063518",
    journal = "Phys. Rev. D",
    volume = "111",
    number = "6",
    pages = "063518",
    year = "2025"
}

@article{Fang:2025vhi,
    author = "Fang, Cheng-Jun and Lyu, Zhen-Hong and Chen, Chao and Guo, Zong-Kuan",
    title = "{Incorporating backreaction in one-loop corrections in ultraslow-roll inflation}",
    eprint = "2502.09555",
    archivePrefix = "arXiv",
    primaryClass = "gr-qc",
    doi = "10.1103/nkrq-3d39",
    journal = "Phys. Rev. D",
    volume = "112",
    number = "2",
    pages = "023547",
    year = "2025"
}

@article{Inomata:2025pqa,
    author = "Inomata, Keisuke",
    title = "{Role of the counterterms in the conservation of superhorizon curvature perturbations at one loop}",
    eprint = "2502.12112",
    archivePrefix = "arXiv",
    primaryClass = "astro-ph.CO",
    doi = "10.1103/r8bh-s48f",
    journal = "Phys. Rev. D",
    volume = "111",
    number = "12",
    pages = "123517",
    year = "2025"
}

@article{Ballesteros:2024cef,
    author = "Ballesteros, Guillermo and Gamb{\'\i}n Egea, Jes{\'u}s and Riccardi, Flavio",
    title = "{Finite parts of inflationary loops}",
    eprint = "2411.19674",
    archivePrefix = "arXiv",
    primaryClass = "hep-th",
    reportNumber = "IFT UAM-CSIC 24-168",
    doi = "10.1007/JHEP06(2025)098",
    journal = "JHEP",
    volume = "06",
    pages = "098",
    year = "2025"
}

@article{Adshead:2008gk,
    author = "Adshead, Peter and Easther, Richard and Lim, Eugene A.",
    title = "{Cosmology With Many Light Scalar Fields: Stochastic Inflation and Loop Corrections}",
    eprint = "0809.4008",
    archivePrefix = "arXiv",
    primaryClass = "hep-th",
    doi = "10.1103/PhysRevD.79.063504",
    journal = "Phys. Rev. D",
    volume = "79",
    pages = "063504",
    year = "2009"
}

@article{Donnelly:2015hta,
    author = "Donnelly, William and Giddings, Steven B.",
    title = "{Diffeomorphism-invariant observables and their nonlocal algebra}",
    eprint = "1507.07921",
    archivePrefix = "arXiv",
    primaryClass = "hep-th",
    reportNumber = "NSF-KITP-15-133",
    doi = "10.1103/PhysRevD.93.024030",
    journal = "Phys. Rev. D",
    volume = "93",
    number = "2",
    pages = "024030",
    year = "2016",
    note = "[Erratum: Phys.Rev.D 94, 029903 (2016)]"
}

@article{Donnelly:2016rvo,
    author = "Donnelly, William and Giddings, Steven B.",
    title = "{Observables, gravitational dressing, and obstructions to locality and subsystems}",
    eprint = "1607.01025",
    archivePrefix = "arXiv",
    primaryClass = "hep-th",
    doi = "10.1103/PhysRevD.94.104038",
    journal = "Phys. Rev. D",
    volume = "94",
    number = "10",
    pages = "104038",
    year = "2016"
}

@article{Giddings:2026fhw,
    author = "Giddings, Steven B. and Wang, Zi-Yue",
    title = "{Leading gravitational dressing of operators and states in de Sitter space}",
    eprint = "2610.02209",
    archivePrefix = "arXiv",
    primaryClass = "hep-th",
    month = "10",
    year = "2026"
}

@article{DeWitt1967,
  author  = {DeWitt, Bryce S.},
  title   = {Quantum Theory of Gravity. I. The Canonical Theory},
  journal = {Physical Review},
  volume  = {160},
  number  = {5},
  pages   = {1113--1148},
  year    = {1967},
  doi     = {10.1103/PhysRev.160.1113}
}

@article{Braglia:2026fzp,
    author = "Braglia, Matteo and Franciolini, Gabriele and Ijaz, Nadir and Peloso, Marco",
    title = "{Mean-Field Backreaction in Single-Field Inflation}",
    eprint = "2610.26669",
    archivePrefix = "arXiv",
    primaryClass = "astro-ph.CO",
    reportNumber = "CERN-TH-2026-239",
    month = "10",
    year = "2026"
}

\end{document}